\documentclass[11pt,openany,oneside]{article}
\usepackage[papersize={210mm,297mm},hdivide={20mm,*,20mm},vdivide={25mm,*,25mm},dvips,mag=1000]{geometry}

\def\bSig\mathbf{\Sigma}

\usepackage[T1]{fontenc}
\usepackage{parskip}
\usepackage{latexsym,amsmath,amssymb, bm, caption, subcaption, bbding, amsthm}
\usepackage{graphicx}
\usepackage{times}
\usepackage{zi4}
\usepackage[toc,page]{appendix}
\usepackage{multirow, multicol}
\usepackage{xcolor, colortbl}
\usepackage{tabularray}
\usepackage{bbm}
\usepackage{pdfpages}
\usepackage{float}
\usepackage{epstopdf}
\usepackage{longtable}
\usepackage{algorithm2e}
\usepackage{soul}
\usepackage{changepage}
\usepackage{natbib}
\usepackage{hyperref}
\hypersetup{colorlinks,allcolors=black}
\usepackage[figuresright]{rotating}

\theoremstyle{definition}

\newcommand\summaryname{Abstract}
\newenvironment{Abstract}%
    {\small\begin{center}%
    \bfseries{\summaryname} \end{center}}

\begin{document}
\setlength{\parindent}{0mm}
\begin{center}
\Large
\newfont{\cmd}{cmdunh10 scaled \magstep3}
\vspace*{-20mm} \hspace*{-0,5cm}
\thispagestyle{empty} 

\vspace{1cm}  
{\LARGE\textbf{Partial Ordering Bayesian Logistic Regression Model for Phase I Combination Trials and Computationally Efficient Approach to Operational Prior Specification}}

Weishi Chen$^1$ and Pavel Mozgunov$^2$\\
$^1$ \textit{weishi.chen@mrc-bsu.cam.ac.uk}\\
$^2$ \textit{pavel.mozgunov@mrc-bsu.cam.ac.uk}\\
$^{1,2} $\textit{MRC Biostatistics Unit, University of Cambridge, Cambridge, UK}
\vspace{1cm}
\end{center}

\begin{Abstract}
\begin{changemargin}{1cm}{1cm}
Recent years have seen increased interest in combining drug agents and/or schedules. Several methods for Phase I combination-escalation trials are proposed, among which, the partial ordering continual reassessment method (POCRM) gained great attention for its simplicity and good operational characteristics. However, the one-parameter nature of the POCRM makes it  restrictive in more complicated settings such as the inclusion of a control group. This paper proposes a Bayesian partial ordering logistic model (POBLRM), which combines partial ordering and the more flexible (than CRM) two-parameter logistic model. Simulation studies show that the POBLRM performs similarly as the POCRM in non-randomised settings. When patients are randomised between the experimental dose-combinations and a control, performance is drastically improved.

Most designs require specifying hyper-parameters, often chosen from statistical considerations (operational prior). The conventional ``grid search'' calibration approach requires large simulations, which are computationally costly. A novel ``cyclic calibration" has been proposed to reduce the computation from multiplicative to additive. Furthermore, calibration processes should consider wide ranges of scenarios of true toxicity probabilities to avoid bias. A method to reduce scenarios based on scenario-complexities is suggested. This can reduce the computation by more than 500 folds while remaining operational characteristics similar to the grid search.
\end{changemargin}
\end{Abstract}

%




%

\section{Introduction}
Phase I trials of new drugs are studies that identify the range of safe doses for further studies. In oncology, a \textit{monotone} relationship between the efficacy and toxicity of the drug is often assumed, and both elements increases with the dose level. Hence, the primary aim of phase I clinical trials is to find the \textit{maximum tolerated dose} (MTD), which is the highest dose level within a \textit{target toxicity level} (TTL). It is shown that good phase I designs (i.e., one that selects the correct MTD with high probability and controls over-dose), model-based designs in particular, can lead to higher success rate in later phases~\citep{Conaway2019PhaseIHelpPhaseIII}. 

For single-agent phase I dose-escalation trials, the most recognised model-based design has been the continual reassessment model (CRM)~\citep{OQuigley1990CRM}. Comparing to rule-based designs, such as the 3+3 design~\citep{Storer1989}, the CRM is advantageous due to its flexibility and efficient use of data~\citep{Hansen2014ModelBasedBetter}. Nevertheless, it also receives criticisms among which, the biggest concern is that one-parameter models are too restrictive for practical uses~\citep{Neuenschwander2008LCRM}. Hence, it was proposed to extend the one-parameter model to more flexible models, such as the two-parameter logistic regression model (BLRM)~\citep{Whitehead1998LogisticCRM, Neuenschwander2008LCRM}. 

In recent years, the study of dose-combinations has attracted increasing clinical interest. The most common case allows the dose of both agent to escalate and de-escalate, and thus the aim is to find the \textit{maximum tolerated combination} (MTC). Extensions of single-agent designs have been proposed, for example, the 2-dimensional BLRM (2BLRM) fits a one-dimensional logistic model for each drug and models the interaction between the them~\citep{Neuenschwander2015,widmer2023principled}. The extension of the original one-parameter CRM, called the partial ordering CRM (POCRM), was proposed by~\citet{Wages2011POCRM}. POCRM focuses on the uncertainty between the ordering of dose-combinations and uses the Bayesian model selection technique to choose the ordering most compatible to data. The POCRM can be adapted to other settings, such as dose-scheduling problems, the combination of more than two drugs, or combination-schedule studies~\citep{Wages2014DoseSchedule,abbas2020comparison,Mozgunov2022PracticalImplementation} and it has been implemented in real trials~\citep{Wages2024POCRMPractical}.

However, since the POCRM uses the one-parameter model as in the CRM, it inherits the criticism of being overly restrictive on the relationship between combinations and toxicities. In particular, in dose-combination studies, additional to locating the MTC, some understanding of the combination-toxicity surface is often required as more than one MTC might exist. Furthermore, there is an increasing interest in the dose-escalation trials that combine the experimental drug with the standard of care and to understand better the interaction between the two compounds and the contribution of components into the toxicity, a randomisation between the experimental arm (typically experimental drug plus the standard of care) and the standard of care alone is introduced. The recent example include the trials in oncology and COVID-19~\citep{Jaki2020AGILE,Mozgunov2019EmaxR,ewings2022practical}. While the one-parameter model is sufficient to accurately estimate of the toxicity probability around the MTD only, under the randomised setting, accurate estimates are required at at least two points of the combination-toxicity surface and warrants the use of more flexible models.   

Most phase I designs, single-agent or dual-agent, require a number of design parameters that should be determined prior to the trial. These often include a ``toxicity skeleton", i.e. a prior estimate of the toxicity probability, at each dose or dose-combination~\citep{LeeCheung2009Skeleton}. For Bayesian approaches, hyperparameters of prior distributions also needs to be specified prior to the trial start. Hence, as the setting becomes complicated and the model becomes flexible, the number of design parameters to be determined inevitably increases. Various calibration approaches have been proposed. \citet{Cheung2002Sensitivity}~considered the ``consistency" of the CRM, i.e. the probability of selecting the correct MTD converges to 1 as the sample size goes to infinity~\citep{Shen1996Consistency} and suggest guessing initial toxicity probabilities based on model sensitivity. \citet{LeeCheung2009Skeleton}~further developed this idea and provide a way to automatically calibrate the toxicity skeletons. Given that the CRM is consistent, \citet{Braun2020SimFreeCalibration}~also suggested a simulation-free way to calibrate design parameters that is not restricted to the toxicity skeleton. An alternative method, the ``grid search" approach, which specifies a grid of potential values for each parameter, and conduct simulations under each value of all parameters, is both systematic and versatile~\citep{Mozgunov2021Interaction, Jaki2020AGILE} and can be adapted to any design. However, the computational cost can be very high and grows multiplicatively with the number of parameters. 

Furthermore, during the calibration process, the designs are evaluated against scenarios of true toxicity probabilities. This leads to a trade-off between the comprehension of the scenarios and the computational cost~\citep{Mozgunov2021Interaction}. Choosing too few scenarios could make the design biased towards the chosen ones, whereas, choosing too many would make the computation expensive. In general, the chosen scenarios should include both extreme cases where all the combinations are way below or beyond the TTL, as well as scenarios that reflect clinical interests~\citep{Wheeler2019Scenarios}. However, the question of the formal assessment of the scenarios to be included in the calibration procedure has not been previously assess in the literature.

The objective of this manuscript is two-fold. Firstly, an extension of the POCRM to a more flexible two-parameter logistic model and a randomised dose-escalation setting is proposed. The proposed approach, referred to as the \textit{partial ordering Bayesian logistic regression model} (POBLRM), adopts the Bayesian model selection technique from POCRM for ordering selection, but is more flexible and is suitable for more complex setting such as the use of a control group. We show that the POBLRM is as accurate, on average, as the POCRM in a non-randomised setting but can drastically improve operating characteristics in randomised settings. Secondly, we present a novel, computationally efficient approach to the calibration of design parameters and the selection of scenarios for the calibration. We propose a novel ``cyclic calibration" method which defines a grid for each hyperparameter but updates the parameter one-at-a-time. Moreover, only a small subset of scenarios are used in simulations, chosen based on scenario complexities. Combining these, the computation cost of the calibration process can be reduced by more than 500 folds while still choosing an operational parameter that has equally good operational characteristics.

\section{Motivating Examples}
\label{Subsec: motivation example} 
This work is motivated by two phase I clinical trials for which the authors contributed as statistical collaborators. The first study is the dose-escalation trial in oncology described by~\citet{Mozgunov2022PracticalImplementation}. The trial studies the combination of an approved drug with an experimental drug. Both of agents can escalated/de-escalated in the trial. Further complexity is coming from the fact that the schedule of administration might also vary. Hence, this presents a 3-dimensional combination-schedule dose-escalation problem with a lot of potential combination-schedules that can be experimented on during the study. The primary objective of the trial is to find the maximum tolerated combination-schedule corresponding to the target risk of the toxicity. The POCRM was agreed to be used in the study as described by~\citet{Mozgunov2022PracticalImplementation}. However, as there are 3 components contributing to the toxicity, a potential secondary objective is to better characterise the combination-schedule toxicity relationship via using a more flexible (than a one-parameter) model. 

The second motivating example is a candidate-specific trial in the AGILE platform, an early phase National UK platform for novel treatments of COVID-19~\citep{Griffiths2021AGILE}. This is a Phase I trial to determine the optimal dose, safety and efficacy of Nitazoxanide~\citet{walker2022open}. Both dose and schedule of Nitazoxanide can vary, implying uncertainties of the dose-schedule toxicity relationship. At stage 1 of the trial, the objective was to determine the safety, tolerability, optimum dose and dosing schedule of Nitazoxanide~\citep{walker2022open} in healthy subjects. The POCRM design was used. At stage 2, however, the objective was to study Nitazoxanide in COVID-19 patients and compare the toxicity to the standard of care (SoC). This is achieved via randomisation of patients between the experimental arm and the SoC (control) arm. Specifically, each cohort of patients are randomised between the SoC and the current MTD. Under this setting, the model should be flexible enough to at least provide accurate fit at the MTD as well as the SoC. Hence, one-parameter models, such as the POCRM, cannot achieve this aim and more flexible models would be preferred.

\section{Methodology}
\label{Sec: POBLRM}
This section starts with an introduction of the POCRM in Section~\ref{Subsec: model design}, which serves as a foundation of the novel POBLRM. Two prior distributions are discussed, the conventional normal prior and a pseudo prior. A link between them is established in Section~\ref{Subsec: prior choice} based on which, a more computationally efficient ordering selection approach is given in Section~\ref{Subsec: ordering selection}.

\subsection{Bayesian Partial Order Continual Reassessment Method (POCRM)}
\label{Subsec: POCRM}

Consider two treatments $A$ and $B$, each with dose levels $a_1<a_2<\cdots<a_r$, $b_1<b_2<\cdots<b_c$, respectively. The combinations can be arranged into a matrix as shown in Table~\ref{Tab: 2x2 grid} for the $2\times2$ combination grid, and the dose-combinations are denoted by $d_i$, $i=1,\ldots,k$, $k=rc$, and indexed by rows, i.e. $d_1=(a_1,b_1), d_2=(a_2,b_1),\ldots$. This paper focuses on settings where, for each single agent, the toxicity probability increases with the dose level, referred to as the \textit{monotonicity} assumption. Nevertheless, the ordering between dose-combinations still contains uncertainties. For example, in the 2-by-2 combination grid, Table~\ref{Tab: 2x2 grid}, $d_3=(a_1,b_2)$ has lower level on drug A but higher level on drug B than $d_2=(a_2,b_1)$. Hence, for subsets of the dose-combinations whose toxicity can be ordered from monotonicity on each single agent, their orderings is called \textit{partial orderings}. Then, each ordering of the whole $k$ combinations respecting partial orderings is called a \textit{complete ordering.} In the $2\times2$ example, monotonicity on drug A and B results partial orderings $d_1\to d_2\to d_4$ and $d_1\to d_3\to d_4$, resulting in 2 possible \textit{complete orderings}, $d_1\to d_2\to d_3\to d_4$ and $d_1\to d_3\to d_2\to d_4$.

\begin{table}[H]
    \centering
    \caption{$2\times 2$ dose-combination grid.\label{Tab: 2x2 grid}}
        \begin{tabular}{l c c}
            \hline
            & \multicolumn{2}{c}{Drug A}\\ \cline{2-3}
            Drug B & $d_3=(a_1,b_2)$ & $d_4=(a_2,b_2)$\\
            & $d_1=(a_1,b_1)$ & $d_2=(a_2,b_1)$\\ \hline
            \end{tabular}
\end{table}

Let $\theta_0$ be the TTL, and $\psi(d_i,a)$ be the toxicity probability at $d_i$, where $a\in\mathcal{A}$ is the model parameter. The \textit{toxicity skeletons} $(\alpha_i)_{i=1}^k$ are prior estimates of the toxicity probability at doses $(d_i)_{i=1}^k$. Assume the toxicity endpoint is binary, with 1 denoting a dose-limiting toxicity (DLT) and 0 a non-DLT. Patients are enrolled in cohorts of $m$. Let $Y_n$ be the number of toxicities in the $n=1,...,N$th cohort, and $x_n$ be the dose level assigned to cohort $n$. Let $\Omega_n$ denote the information set after enrolling cohort $n$, i.e. $\Omega_n=\{x_1,y_1,...,x_{n}, y_{n}\}$. The POCRM design has two steps, ordering selection and combination selection. 
\begin{itemize}
    \item[(1)] Ordering selection.\\
        Upon enrolling the $n$th cohort, the posterior probability of ordering $s$, $s=1,...,S$, is calculated as $p(s\,|\,\Omega_n)\propto \mathbb{P}_s(\Omega_n)p(s)$ where, $p(s)$ is the prior probability of ordering $s$. $\mathbb{P}_s(\Omega_n)$ denotes the marginal likelihood under ordering $s$, $\mathbb{P}_s(\Omega_n)=\int_\mathcal{A}\pi(a)\mathcal{L}_s(a\,|\,\Omega_n)d\theta$, 
        where $\pi(a)$ denotes the prior distribution of the model parameter $a$, and $\mathcal{L}_s$ is the likelihood under ordering $s$. The ordering with the highest posterior probability, $s^\ast$, is selected.
    \item[(2)] Combination selection. Applying the CRM design.\\
        Under the selected ordering $s^\ast$, let $\pi_{s^\ast}(\cdot|\Omega_n):\mathcal{A}\mapsto[0,1]$ be the current belief before enrolling the $n$th cohort. Then, after enrolling cohort $n$, the updated belief becomes $\pi_{s^\ast}(a\,|\,\Omega_{n+1})\propto\pi_{s^\ast}(a|\Omega_n)\mathcal{L}_{s^\ast}(a;x_n,y_n)$, where $\mathcal{L}_{s^\ast}(a;x_{n},y_n)=\mathrm{Bin}(\,y_n;m,\psi_{s^\ast}(x_n,a)\,)$. The CRM design uses a one-parameter model $\psi_{s}(d_i,a)=\alpha_{s,i}^{a}$, where, $\alpha_i$ is the toxicity skeleton, and $(\alpha_{s,i})_{i=1}^k$ is a permutation of $(\alpha_i)_{i=1}^k$ according to ordering $s$. The posterior mean $\hat{a}$ is plugged into the model to estimate the toxicity probability at $d_i$, $\hat{p}_i=\alpha_i^{\hat{a}}$. The $(n+1)$th cohort will be allocated to dose level $x_{n+1}$ whose estimated toxicity is closest to $\theta_0$.
\end{itemize}
Iterating steps (1), (2), the recommended MTC will be $x_{N+1}$ after all patients are exhausted.

\subsection{The POBLRM model}
\label{Subsec: model design}
Let $\psi_s(d_i,\theta)$ be the new working model under ordering $s$, where $\theta=(\theta_1,\theta_2)^T$ is a vector of model parameters. The two-parameter logistic model is parameterised as
\[\mathrm{logit}\left(\psi_s(d_i,\theta)\right)=\log\left(\frac{\psi_s(d_i,\theta)}{1-\psi_s(d_i,\theta)}\right)=\theta_1+\theta_2\tilde{d}_i,\quad i=1,...,k,\,\,s=1,...,S.\]
\textit{Standardised dose level} $\tilde{d}_i$'s are used, which are elicited from prior estimates of toxicity probabilities (i.e. the skeleton) at each combination. Denote the prior toxicity at $d_i$ by $\hat{p}_i^{(0)}$, and let $\hat{\theta}^{(0)}=\left(\hat{\theta}_1^{(0)}, \hat{\theta}_2^{(0)}\right)^T$ be the prior point estimates of model parameters, then the prior beliefs should be compatible, i.e.
$\mathrm{logit}\left(\hat{p}_i^{(0)}\right)=\hat{\theta}_1^{(0)}+\hat{\theta}_2^{(0)}\tilde{d}_i$, which gives $\tilde{d}_i=\left[\mathrm{logit}\left(\hat{p}_i^{(0)}\right)-\hat{\theta}_1^{(0)}\right]/\hat{\theta}_2^{(0)}$, $i=1,...,k$. In the rest of the paper, we will abuse the notation $d_i$ to denote both the dose combination and the standardised dose levels. Under the Bayesian approach, a prior distribution $\pi(\theta)$ is assigned to the two-dimensional model parameter $\theta$.

Adaptations of the POBLRM should be made under the randomised setting, where a control arm is introduced and patients are randomised between the SoC and dose-combinations. Denote the standardised dose level at SoC $d_0$, then, the logistic model is formed in the same way as the non-randomised setting with an additional constraint $d_0=0$. This is to ensure updates on the distribution of $\theta_1$ based on information from the treatment arm will not affect the estimate at the SoC. The prior mean $\mu_1$, whence, corresponds to the prior estimate of the toxicity at the SoC. The rest of the model is not altered under the randomised setting.

\subsection{Prior distributions}
\label{Subsec: prior choice}

The bivariate normal prior is commonly used for BLRM, 
$
(\theta_1,\log\theta_2)^T\sim\mathcal{N}(\mu,\Sigma)
$,
where $\mu=\begin{pmatrix}
\mu_1\\
\mu_2
\end{pmatrix}$ and 
$\Sigma=\begin{pmatrix}
\sigma_1^2 & 0\\
0 & \sigma_2^2
\end{pmatrix}$. However, either numeric double integration or MCMC is required for model estimation, which makes BLRM more computationally expensive compared to CRM. This problem amplifies when the BLRM should be run under each of the potential $S$ orderings. This motivates the use of conjugate priors, which leads to simple calculations and easy interpretations of the hyperparameters~\citep{Whitehead1998LogisticCRM}.

Before starting the trial, assume two pseudo cohorts of patients with sample sizes $n_{-1}$ and $n_0$ are enrolled to the lowest and highest combinations $d_1$ and $d_k$, result in $y_{-1}$ and $y_0$ DLTs, respectively. Denote $u_j=n_j-y_j$, $j=-1,0$. Since these experiments are not performed on real patients, none of the numbers needs to be integer. Let $p_{-1}=\psi(d_1,\theta)$ and $p_0=\psi(d_k,\theta)$ be the toxicity probabilities at those two combinations. Then, independent beta priors are defined on $p_{-1}$ and $p_0$, $p_{j}\stackrel{indep.}{\sim}\mathrm{Beta}(y_j,u_j)$, $j=-1, 0$. This implies a prior on $\theta$, $\pi(\theta)\propto \prod_{j=-1}^0 p_j^{y_j}(1-p_j)^{u_j}$.
After $n$ cohorts, suppose the selected ordering is $s^\ast$, and let $p_j=\psi_{s^\ast}(x_j,\theta)$, $u_j=m-y_j$, the likelihood and posterior of $\theta$ takes the form in~\eqref{Eqn: pseudo likelihood} and~\eqref{Eqn: pseudo posterior}, respectively. The only difference is $j$ starts from -1 rather than 1. Thus, the prior acts as if two extra pseudo experiments are performed, whence the name \textit{pseudo prior}.
\begin{align}
    \mathcal{L}(\theta\,|\,\Omega_n)\propto&\prod_{j=0}^{n}p_{j}^{y_j}(1-p_{j})^{u_j},\label{Eqn: pseudo likelihood}\\
    \pi(\theta\,|\,\Omega_{n+1})\propto&\prod_{j=-1}^{n}p_{j}^{y_j}(1-p_{j})^{u_j}\,\,.
    \label{Eqn: pseudo posterior}
\end{align}

Besides the convenience on interpretation, the pseudo prior also allows a frequentist view and selects ordering based on the AIC. Hence, the double integrals required from the marginal likelihood can be avoided, which can lead to significant reduction on computation time.

Although the pseudo prior is advantageous in interpretation and computation, normal priors are more conventional~\citep{Jaki2020AGILE}. This motivates establishing a link between them. Given \textit{normal hyperparameters} $(\mu_1,\mu_2,\sigma_1,\sigma_2)$, we propose a way to find the \textit{pseudo hyperparameters} $(y_{-1},n_{-1},y_0,n_0)$ such that the pseudo prior best matches the given normal prior. Then, the pseudo prior can be used for simulations to reduce the computational time.

Let $\pi_\mathcal{N}(\theta\,|\,\Omega_n)$ and $\pi_\beta(\theta\,|\,\Omega_n)$ be the posterior of $\theta$ under normal and pseudo prior before enrolling the $(n+1)$th cohort, and let $\psi_\mathcal{N}(d_i,\theta\,|\,\Omega_n)$ and $\psi_\beta(d_i,\theta\,|\,\Omega_n)$ be posterior toxicity probabilities at dose $d_i$ under normal and pseudo prior, respectively. The strategy is to match  the priors ($n=0$). The KL-divergence metric of mismatch is proposed to be used which is expressed as $\mathbb{D}_\mathrm{KL}(\pi_\mathcal{N}\|\pi_\beta)=\mathbb{E}_{\theta\sim\pi_\mathcal{N}(\theta\,|\,\Omega_n)}\left[\log\frac{\pi_\mathcal{N}(\theta\,|\,\Omega_n)}{\pi_\beta(\theta\,|\,\Omega_n)}\right]$.

The comparison of the KL-divergence metric to two other metrics are given in Supplementary Materials Section~1. It shows that the proposed metric is the most stable one and gives the highest proportion of matching recommendations. Moreover, it has been demonstrated that using the KL-divergence prior matching and the AIC criterion leads to approximately the same operating characteristics while reducing the computation time by nearly 3-fold in the $2\times3$ dose-combination and 20-fold in the $3\times3$ case.

\subsection{POBLRM Design}
\label{Subsec: ordering selection}
Hence, with the ordering selection based on AIC, the POBLRM design is summarised in Algorithm~\ref{Alg: POBLRM}. The initialisation part defines the standardised dose levels and match the normal prior to pseudo prior. Then, for each newly enrolled cohort $n$, iterate between two steps: ordering selection and combination selection. The ordering selection step fits a GLM treating the pseudo prior as two more cohorts, and select the ordering with the smallest AIC. Then, the combination selection step fits a BLRM under the selected ordering and recommends the combination with toxicity closest to the TTL. The distinction between randomised and non-randomised versions is that, if randomised, $m_2$ patients in each cohort are randomised to the SoC $d_0$ and the remaining $m_1$ patients are assigned to the current MTC. Whereas, under the non-randomised version, all patients are allocated to the MTC.

\section{Cyclic Calibration Approach}\label{Sec: calibration}
\subsection{Calibration setting}
\label{Subsec: cyclic calibration}
In the POBLRM design, a set of design parameters needs elicitation. This can either come from clinicians if they have good understandings of the treatments, or often, when prior opinions are vague, \textit{operating parameters} are used. That is, the design parameters giving the best performance across various scenarios, measured by the PCS across a comprehensive set of scenarios~\citep{Mozgunov2021Interaction}. To reduce the number of design parameters, equal-spaced toxicity skeleton is used, $\hat{p}_i^{(0)}=\hat{p}_1^{(0)} + (i-1)\nu$, $i=2,\ldots,k$. Then, for POBLRM, the design parameters are (i) the prior toxicity probability at $d_1$, $\hat{p}_1^{(0)}$; (ii) the space between prior toxicities $\nu$; (iii) the mean and variance of the normal prior $\mu_1,\mu_2, \sigma_1^2,\sigma_2^2$. 

A grid of values has been defined for each parameter to choose from. Let $\hat{p}_1^{(0)}\in\mathfrak{P}$, $\nu\in\mathfrak{V}$, $\mu_1,\mu_2\in\mathfrak{M}$, and $\sigma_1,\sigma_2\in\mathfrak{S}$. Denote $\omega=\{\hat{p}_1^{(0)},\nu,\mu_1,\mu_2,\sigma_1,\sigma_2\}$. The mean PCS is used as a metric to assess the OC of each potential parameter value $\omega$. Explicitly, among a set of scenarios $c=1,...,C$ of true toxicity probabilities, estimate the corresponding PCS under the parameter value $\omega$ as $\mathrm{PCS}^{(c)}(\omega)$. Then, calculate the geometric mean of the PCS
\begin{equation}
    \mathrm{PCS}(\omega)=\left(\prod_{c=1}^C\mathrm{PCS}^{(c)}(\omega)\right)^{1/C},
    \label{Eqn: geometric mean}
\end{equation} 
and pick the set of parameter values that maximise the mean PCS.

The conventional method to calibrate the parameters is the ``grid search", which considers all possible values of $\omega$ in the set $\mathfrak{P}\times\mathfrak{V}\times\mathfrak{M}^2\times\mathfrak{S}^2$~\citep{Jaki2020AGILE}. Then, the value of $\omega$ leading to the maximum $\mathrm{PCS}(\omega)$ would be selected as an operational prior. The computational cost could be very high. For example, even if all the grids only contain 5 values, which is a moderate size, it mounts to $5^6=15625$ potential values of $\omega$. Then, doing $10^4$ simulations for each $\omega$ under each scenario requires roughly $10^9$ simulations for $C=10$.

\subsection{Proposed approach}
\label{Subsec: cyclic algorithm}
Alternatively, the parameters could be updated one-at-a-time. Starting with some random $\omega$, values of $\hat{p}_1^{(0)}\in\mathfrak{P}$ is considered, while holding the value of the other 5 parameters fixed. The $\hat{p}_1^{(0)}$ that maximises the PCS will be selected and fixed, and then values of $\nu\in\mathfrak{V}$ will be considered, so on and so forth. Since the algorithm cycles through parameters, this is named the \textit{cyclic calibration} approach. When all six parameters are considered once, this is called one \textit{cycle}. The algorithm stops once staying at the same $\omega$ for a cycle.

\RestyleAlgo{ruled}
\begin{algorithm}[H]
\footnotesize
\SetKwInOut{Input}{Input}\SetKwInOut{Output}{Output}
\caption{Partial Order Bayesian Logistic Regression Model (POBLRM)}
\Input {Prior toxicity probabilities $\hat{p}^{(0)}$\; 
Parameters of the normal prior $\mu_1,\mu_2,\sigma_1,\sigma_2$\;
Complete orderings to be included $s=1,...,S$\;
Prior probability of orderings $\{p(s)\}_{s=1}^S$\; 
Sample size $N\times m$, where $N$ is the number of cohorts\;
Cohort size $m=m_1+m_2$, where $m_1,m_2$ are the number of patients allocated to the dose-combinations and SoC\; }
\Output {the recommended MTC\;}
\Begin(Initialisation){
Calculate the prior point estimates of model parameters
$\hat{\theta}_1^{(0)}=\mu_1$, $\hat{\theta}_2^{(0)}=\exp\left(\mu_2+\frac{\sigma_2^2}{2}\right)$\;
Calculate the standardised dose levels
$d_i=\frac{\mathrm{logit}\left(\hat{p}_i^{(0)}\right)-\hat{\theta}_1^{(0)}}{\hat{\theta}_2^{(0)}}$, $i=1,...,k$\;
Match the normal prior $\mathcal{N}(\mu,\Sigma)$ to the pseudo prior with hyperparameters $(y_{-1}, n_{-1}, y_0, n_0)$ via KL-divergence\;
	\uIf{randomised}{
	Let $d_0=0$, $\hat{p}_0^{(0)}=\mu_1$\;
	Randomise $m_1$ patients to $d_1$ and $m_2$ patients to $d_0$\;
	}
	\Else {
	Enrol the first cohort of $m$ patients to $d_1$\;
	}
}
\BlankLine

\For{$n\gets 1$ \KwTo $N$} {
	\Begin(Ordering selection){
	\For{$s\gets 1$ \KwTo $S$} {
		Fit logistic GLM to the dataset $\{(y_{-1}, n_{-1}), (y_0,n_0),\Omega_n\}$ under ordering $s$\;
		Extract the AIC from the GLM output\;
	}
	Select the ordering $s^\ast$ with the smallest AIC\;
	}
	\Begin(Combination selection){
	Under $s^\ast$, calculate the posterior distribution of $\theta$,
	$\pi(\theta\,|\,\Omega_n)\propto\pi(\theta)\prod_{j=1}^n\mathrm{Bin}(y_j;m,\psi_{s^\ast}(x_j,\theta))$, where $\psi$ is the two-dimensional BLRM\;
	Calculate the posterior mean of $\theta$ as $\hat{\theta}=\mathbb{E}[\theta\,|\,\Omega_n]$\;
	Update the estimated toxicity probability at each combination as
	$\hat{p}_i=\psi_{s^\ast}(d_i, \hat{\theta})$\;
	Select the dose-combination to enrol the next cohort
	$\left|\psi_{s^\ast}(x_{n+1},\hat{\theta})-\theta_0\right|=\min_i|\hat{p}_i-\theta_0|$\;
	}
	\uIf{randomised}{
	Randomise $m_1$ patients to $x_{n+1}$ and $m_2$ patients to $d_0$\;
	}
	\Else {
	Enrol the next cohort of $m$ patients to $x_{n+1}$\;
	}
  }
The final recommended MTC will be $x_{N+1}$.
\label{Alg: POBLRM}
\end{algorithm}

In practice, the PCS is estimated by simulating the trial $10^4$ times and calculating the proportion of times the true MTC has been selected. Hence, the PCS contains estimation error. When seeing an improvement in the PCS from some parameter value $\omega$ to $\omega'$, a natural question to ask is whether this improvement indicates superiority of $\omega'$ or it is simply due to randomness. One solution is to build a confidence interval (C.I.) around each point estimate of the PCS. Let $K$ be the number of times that the correct MTC has been selected under $B$ simulations, then $K\sim\mathrm{Bin}(B, \mathrm{PCS}_0)$, where $\mathrm{PCS}_0$ denotes the true (unknown) PCS. The central limit theorem suggests that $\sqrt{B}\left(\frac{K}{B}-\mathrm{PCS}_0\right)\stackrel{d}{\to}\mathcal{N}(0,\mathrm{PCS}_0(1-\mathrm{PCS}_0))$.
Hence, a $100(1-\alpha)\%$ C.I. for $\mathrm{PCS}_0$ would be $\mathrm{PCS}\pm z_{1-\alpha/2}\frac{1}{\sqrt{B}}\mathrm{PCS}(1-\mathrm{PCS})$, 
where $\mathrm{PCS}=K/B$ is the estimated PCS, and $z_{\alpha}$ denotes the $\alpha$ quantile of $\mathcal{N}(0,1)$.

Comparing two sets of design parameters $\omega,\omega'$, we consider an improvement from $\omega$ to $\omega'$ only if the upper bound of $\mathrm{PCS}(\omega)$ is smaller than the lower bound of $\mathrm{PCS}(\omega')$. The confidence level $\alpha$ should be chosen such that it is neither too large to deny any improvement nor too small to make the computational cost too high. This is summarised in Algorithm \ref{Alg: cyclic}.

\RestyleAlgo{ruled}
\begin{algorithm}[!htb]
\SetKwInOut{Input}{Input}\SetKwInOut{Output}{Output}
\caption{Cyclic calibration}
\Input{Grids of design parameters $\mathfrak{P},\mathfrak{V},\mathfrak{M},\mathfrak{S}$, Confidence level $\alpha$.} 
\Output{ the set of operational parameters $\omega=(\hat{p}_1^{(0)}, \nu,\mu_1,\mu_2,\sigma_1,\sigma_2)$}
\BlankLine

\Begin(Initialisation){
	Randomly generate a starting point $\omega_0\in\mathfrak{P}\times\mathfrak{V}\times\mathfrak{M}^2\times\mathfrak{S}^2$\;
	iteration: $i\gets 0$\;
}
\BlankLine

\While{$i=0$ or $\omega_i\not\in\{\omega_0,\cdots,\omega_{i-1}\}$} {
	$\omega^\ast\gets\omega_i$\;
	\For{$\ell\gets 1$ \KwTo 6}{
		Search over the grid of the $\ell$th parameter, fixing the others\; 
		Let $\tilde{\omega}$ be the one with the best PCS over this search\;
		Construct $(1-\alpha)$ C.I. for $\mathrm{PCS}(\omega^\ast)$, $(\mathrm{PCS}\left(\omega^\ast)^L,\mathrm{PCS}(\omega^\ast)^U\right)$, and for $\mathrm{PCS}(\tilde{\omega})$, $\left(\mathrm{PCS}(\tilde{\omega})^L,\mathrm{PCS}(\tilde{\omega})^U\right)$\;
		\uIf{$\mathrm{PCS}(\omega^\ast)^U<\mathrm{PCS}(\tilde{\omega})^L$}{
			$\omega^\ast\gets\tilde{\omega}$		
		} 
		\Else {$\omega^\ast\gets\omega^\ast$}
	}
	$\omega_{i+1}\gets\omega^\ast$\;
	$i\gets i+1$\;
  }
The operational parameter would be the final $\omega_i$.
\label{Alg: cyclic}
\end{algorithm}

Note that each time $\omega$ is updated, the PCS only goes up, and the same $\omega$ is never visited twice. These guarantee the convergence of algorithm~\ref{Alg: cyclic}, because the number of possible $\omega$ is finite. Moreover, the worst case is to exhaust the set $\mathfrak{P}\times\mathfrak{V}\times\mathfrak{M}^2\times\mathfrak{S}^2$, i.e. the grid search. Thus, the computation is guaranteed to be lower than, or at most equals to, the grid search.

To choose the confidence level $\alpha$, one can define a trade-off function. For example, 
\begin{equation}
    S_{\alpha,\lambda}=\frac{\mathrm{PCS}_\alpha-\mathrm{PCS}_0}{\mathrm{PCS}_0}-\lambda\frac{T_\alpha-T_{\min}}{T_{\min}},
    \label{Eqn: conf level score}
\end{equation}
where $\mathrm{PCS}_\alpha$ is the final PCS under significance level $\alpha$, $\mathrm{PCS}_0$ is the initial PCS at the starting point, $T_\mathrm{\min}$ and $T_\alpha$ are the minimum possible computation time and the computation time under significance level $\alpha$, respectively. The computation time is included as the number of parameter combinations searched over. The tuning parameter $\lambda$ controls the trade-off, large values sacrifice accuracy for computation. For a fixed $\lambda$, the significance level $\alpha$ that maximises the score is selected. Table~4 in Supplementary Materials provides an example where the score under 12 values of $\alpha$ and 4 different $\lambda$ are listed.  

\subsection{Choice of calibration scenarios}
\label{Subsec: subset calibration}
Since the true toxicity probabilities are unknown in practice, a phase I design should be able to accommodate all possible combination-toxicity relationships. Specifically, in dose-escalation trials, this can be redefined as ``all possible (combination of) locations of the MTC in the combination grid''. We refer to a combination-toxicity relationship implying certain locations of the MTCs as  a ``scenario''. For any calibration procedure, one has to specify the set of scenarios that will be used. When choosing such scenarios, two elements should be considered: (i) the position of the true MTC, and (ii) the ``difficulty'' of the scenario. 

\begin{figure}[H]
    \centerline{\includegraphics[width=0.45\linewidth]{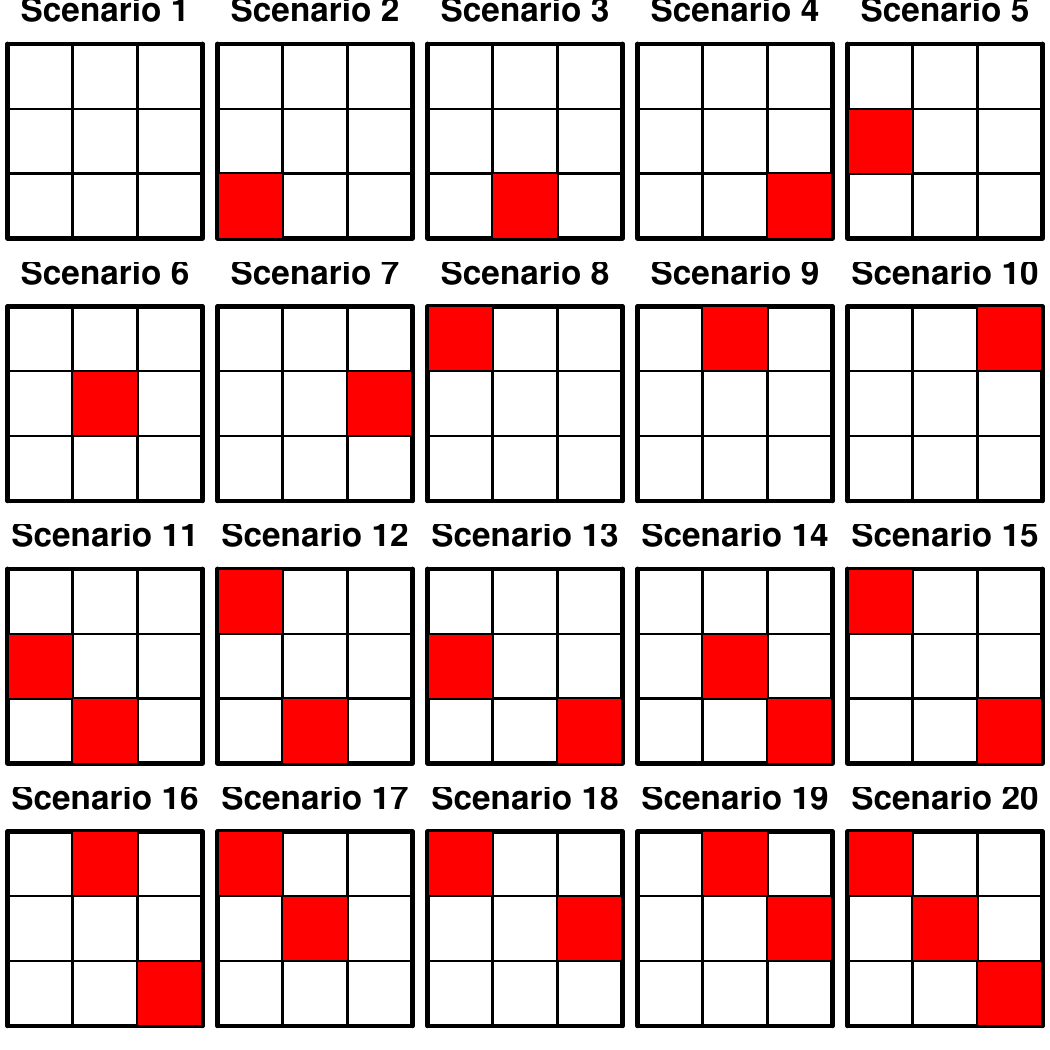}}
    \caption{The full set of scenarios for $3\times 3$ combinations. The position of the MTC in red.}
    \label{Fig: full set 3x3}
\end{figure}

For the former, the scenarios should cover all possible position of the MTC or positions when there are more than one true MTC. This is referred to as the \textit{full set of scenarios}. Taking the $3\times3$ case as an example, the possible position(s) of the MTC(s) are listed as follows and graphically shown in Figure~\ref{Fig: full set 3x3}, which include the scenarios with (i) no MTC (Scenario 1), all combinations are more toxic than the TTL; (ii) one MTC (Scenario 2-10), one for each combination being the MTC; (iii) two MTCs (Scenario 11-19), one for each pair of combinations being the MTC respecting monotonicity; (iv) three MTCs (Scenario 20). 
Given the position of the MTC, the difference in toxicity probability between neighbouring combinations is chosen to be clinically meaningful to detect, commonly $10\%-15\%$. Hence, after setting the toxicity of the MTC to TTL, its neighbours are set to TTL$\pm10\%$. The other combinations will then each have a difference of 10\%, following monotonicity of each single agent. For combinations with ambiguous orderings, their toxicities are set equal. 

Ideally, one should include all possible scenarios to the calibration, which, however, noticeably increases the computation cost of any procedure. Instead, it is argued that a subset of scenarios with the MTCs covering the location at the beginning and ``top'' of the combination grid should be selected~\citep{Wheeler2019Scenarios, Mozgunov2021Interaction}. Possible sets include:
\begin{itemize}
    \item Scenarios 2 and 10 in Figure~\ref{Fig: full set 3x3} with the MTCs at the corners, referred to as ``Pos2''
    \item Scenario 2, 11, 19, 10 in Figure~\ref{Fig: full set 3x3} with the MTCs going from the bottom-left corner to the top-right corner, referred to as ``Pos4''
    \item Scenario 2, 11, 20, 19, 10 in Figure~\ref{Fig: full set 3x3} with the MTCs going from the bottom-left corner to the top-right corner, referred to as ``Pos5''
\end{itemize}

This approach, however, does not explicitly taken into account how challenging it is to identify the MTC. Regarding the ``difficulty'' of a scenario, the partial ordering benchmark (PO-benchmark) proposed by~\citet{Mozgunov2022Benchmark} is adopted as such a measure, which provides an upper bound on the PCS for all dose-combination designs under a given scenario (detailed in Section~4.1 in supplementary materials). We propose to use the PO-becnhmark to select the subset scenarios for calibration. Specifically, the selection is among
\begin{itemize}
    \item Scenario with the lowest and highest PCS under the benchmark, referring to as ``Ben2''
    \item Two scenarios with the lowest PCS and two with the highest PCS under the benchmark, referring to as ``Ben4''
\end{itemize}

To select the approach to the choice of the calibration scenarios, the score in~\eqref{Eqn: scen subset score} is proposed to quantify the balance between the comprehension of scenarios and the computation (details in Section~4.2 of supplementary materials). The subscript $\mathrm{scen}$ refers to subsets of the full set of scenarios, which are chosen based on the PO-benchmark or the positions.
\begin{equation}
    S_{\mathrm{scen},\xi}=\frac{\mathrm{PCS}_\mathrm{scen}-\mathrm{PCS}_0}{\mathrm{PCS}_0}-\xi\frac{T_\mathrm{scen}-T_{\min}}{T_{\min}}.
    \label{Eqn: scen subset score}
\end{equation}

Based on simulation results (Section~4.2 of SM), we calibrating parameters using the cyclic algorithm on the subset of ``Ben2'' scenarios. Compared to the full set of scenarios, Ben2 reduces the computational cost by 10 folds, whereas the difference in the PCS is within 0.5\%.

\section{Numerical evaluation} 
\label{Sec: Simulation}
In this section, we illustrate how the proposed design and calibration approach can be used in practice and compare their performance to competing approaches. The \texttt{R} code to reproduce the results is provided on GitHub \verb|https://github.com/WeishiC/Partial-Ordering-BLRM|. 

\subsection{Setting}
\label{Subsec: non-randomised setting}
Consider a $3\times3$ dual-agent combination-escalation problem with agents, $A$ and $B$, each has 3 dose levels. Monotonicity is assumed within each agent, with the dose of the other being fixed. The objective is to identify the MTC corresponding to the target toxicity probability is $\theta_0=0.3$. The combination grid is given in Table~\ref{Tab: 3x3 grid}. Up to 45 patients are to be recruited and in cohorts of $m=3$ patients. The starting combination is the lowest combination $(a_1,b_1)$. 

\begin{table}[H]
    \centering
    \begin{tabular}{l c c c}
        \hline
        & \multicolumn{3}{c}{Drug A}\\ \cline{2-4}
        Drug B& $a_1$ & $a_2$ & $a_3$\\ \hline
        $b_3$ & $d_7$ & $d_8$ & $d_9$\\
        $b_2$ & $d_4$ & $d_5$ & $d_6$\\
        $b_1$ & $d_1$ & $d_2$ & $d_3$\\ \hline
    \end{tabular}
    \caption{$3\times 3$ dose-combination. \label{Tab: 3x3 grid}}
\end{table}

Similarly to POCRM, POBLRM requires specifying toxicity orderings prior to the trial. We adopt the 6 orderings suggested by~\citet{Wages2013Orders} with equal prior probabilities
\begin{itemize}
    \item By rows: $1\to2\to3\to4\to5\to6\to7\to8\to9$,
    \item By columns: $1\to4\to7\to2\to5\to8\to3\to6\to9$,
    \item Up diagonals: $1\to2\to4\to3\to5\to7\to6\to8\to9$,
    \item Down diagonals: $1\to4\to2\to7\to5\to3\to8\to6\to9$,
    \item Up-and-down diagonals: $1\to2\to4\to7\to5\to3\to6\to8\to9$,
    \item Down-and-up diagonals: $1\to4\to2\to3\to5\to7\to8\to6\to9$.
\end{itemize}

Assuming the TTL of 30\% and the difference between neighbouring combination of 10\%, there are 20 possible combination-toxicity scenarios that cover different location of the MTCs and the different combination of locations. The considered scenarios are given in Table~\ref{Tab: example scenarios}. 

\begin{table}[H]
    \centering
    \begin{tabular}{l ccc c ccc c ccc c ccc}
        \hline
        Drug & \multicolumn{3}{c}{Drug A} && \multicolumn{3}{c}{Drug A} && \multicolumn{3}{c}{Drug A} && \multicolumn{3}{c}{Drug A}\\ \cline{2-4}\cline{6-8}\cline{10-12}\cline{14-16}
        B & $a_1$ & $a_2$ & $a_3$ && $a_1$ & $a_2$ & $a_3$ && $a_1$ & $a_2$ & $a_3$ && $a_1$ & $a_2$ & $a_3$\\ \hline
        & \multicolumn{3}{c}{Scenario 1} && \multicolumn{3}{c}{Scenario 2} && \multicolumn{3}{c}{Scenario 3} && \multicolumn{3}{c}{Scenario 4}\\
        $b_3$& 0.60 & 0.70 & 0.80 && 0.50 & 0.60 & 0.70 && 0.50 & 0.60 & 0.70 && 0.40 & 0.50 & 0.60\\
        $b_2$& 0.50 & 0.60 & 0.70 && 0.40 & 0.50 & 0.65 && 0.40 & 0.50 & 0.60 && 0.20 & 0.40 & 0.50\\
        $b_1$& 0.40 & 0.45 & 0.60 && \textbf{0.30} & 0.40 & 0.50 && 0.20 & \textbf{0.30} & 0.50 && 0.10 & 0.20 & \textbf{0.30}\\\hline

        & \multicolumn{3}{c}{Scenario 5} && \multicolumn{3}{c}{Scenario 6} && \multicolumn{3}{c}{Scenario 7} && \multicolumn{3}{c}{Scenario 8}\\
        $b_3$& 0.50 & 0.60 & 0.70          && 0.40 & 0.50 & 0.60 && 0.40 & 0.50 & 0.60          && \textbf{0.30} & 0.50 & 0.60\\
        $b_2$& \textbf{0.30} & 0.50 & 0.60 && 0.20 & \textbf{0.30}& 0.50 && 0.20 & 0.25 & \textbf{0.30} && 0.10 & 0.20 & 0.40\\
        $b_1$& 0.20          & 0.40 & 0.50 && 0.10 & 0.15 & 0.20 && 0.10 & 0.20 & 0.25          && 0.05 & 0.10 & 0.25\\ \hline

        & \multicolumn{3}{c}{Scenario 9} && \multicolumn{3}{c}{Scenario 10} && \multicolumn{3}{c}{Scenario 11} && \multicolumn{3}{c}{Scenario 12}\\
        $b_3$& 0.20 & \textbf{0.30} & 0.60 && 0.20 & 0.25 & \textbf{0.30} && 0.40 & 0.50 & 0.70          && \textbf{0.30} & 0.50 & 0.60\\
        $b_2$& 0.10 & 0.20 & 0.50 && 0.15 & 0.20 & 0.25          && \textbf{0.30} & 0.40 & 0.60 && 0.20 & 0.40 & 0.50\\
        $b_1$& 0.05 & 0.10 & 0.40 && 0.05 & 0.15 & 0.20          && 0.20 & \textbf{0.30} & 0.50 && 0.10 & \textbf{0.30} & 0.40\\ \hline

        & \multicolumn{3}{c}{Scenario 13} && \multicolumn{3}{c}{Scenario 14} && \multicolumn{3}{c}{Scenario 15} && \multicolumn{3}{c}{Scenario 16}\\
        $b_3$& 0.40 & 0.50 & 0.60          && 0.40 & 0.50 & 0.60          && \textbf{0.30} & 0.50 & 0.60 && 0.20 & \textbf{0.30} & 0.50\\
        $b_2$& \textbf{0.30} & 0.40 & 0.50 && 0.20 & \textbf{0.30} & 0.40 && 0.15 & 0.20 & 0.50 && 0.10 & 0.20 & 0.40\\
        $b_1$& 0.10 & 0.20 & \textbf{0.30} && 0.10 & 0.20 & \textbf{0.30} && 0.05 & 0.15 & \textbf{0.30} && 0.05 & 0.10 & \textbf{0.30}\\ \hline

        & \multicolumn{3}{c}{Scenario 17} && \multicolumn{3}{c}{Scenario 18} && \multicolumn{3}{c}{Scenario 19} && \multicolumn{3}{c}{Scenario 20}\\
        $b_3$& \textbf{0.30} & 0.50 & 0.60 && \textbf{0.30} & 0.50 & 0.60 && 0.25 & \textbf{0.30} & 0.40 && \textbf{0.30} & 0.40 & 0.60\\
        $b_2$& 0.20 & \textbf{0.30} & 0.50 && 0.15 & 0.20 & \textbf{0.30} && 0.15 & 0.25 & \textbf{0.30} && 0.20 & \textbf{0.30} & 0.50\\
        $b_1$& 0.10 & 0.20 & 0.40 && 0.05 & 0.15 & 0.25 && 0.05 & 0.20 & 0.25          && 0.10 & 0.20 & \textbf{0.30}\\ \hline 
    \end{tabular}
    \caption{True toxicity probabilities for $3\times3$ dose-combinations. The TTL is 0.3. \label{Tab: example scenarios}}
\end{table}

When evaluating operational characteristics of the designs, the main consideration is the geometric mean of the PCS across the above 20 scenarios, defined in Equation~\eqref{Eqn: geometric mean}, and the probability of overdosing, i.e. assign patients to a combination with toxicity probability higher than the TTL. The computational time will also be considered as a secondary object.

\subsection{Cyclic calibration of the POBLRM}
Below, the application of the cyclic calibration approach to the POBLRM design is presented. As per Section~4, there six designs parameters to be calibrated. The following grids are used 
\begin{align*}
    &\hat{p}_1^{(0)}\in\mathfrak{P}=\{0.01, 0.05, 0.1, 0.15, 0.2, 0.25, 0.3\}, &&
    \nu\in\mathfrak{V}=\{0.01, 0.05, 0.1, 0.15\},\\
    &\mu_1, \mu_2\in\mathfrak{M}=\{-1, 0, 1, 2, 3, 4\}, &&
    \sigma_1, \sigma_2\in\mathfrak{S}=\{0.5, 1, 2, 5, 10\}.
\end{align*}

\begin{figure}[H]
    \centering
    \begin{subfigure}[b]{0.48\linewidth}
        \centering
        \includegraphics[width=\textwidth]{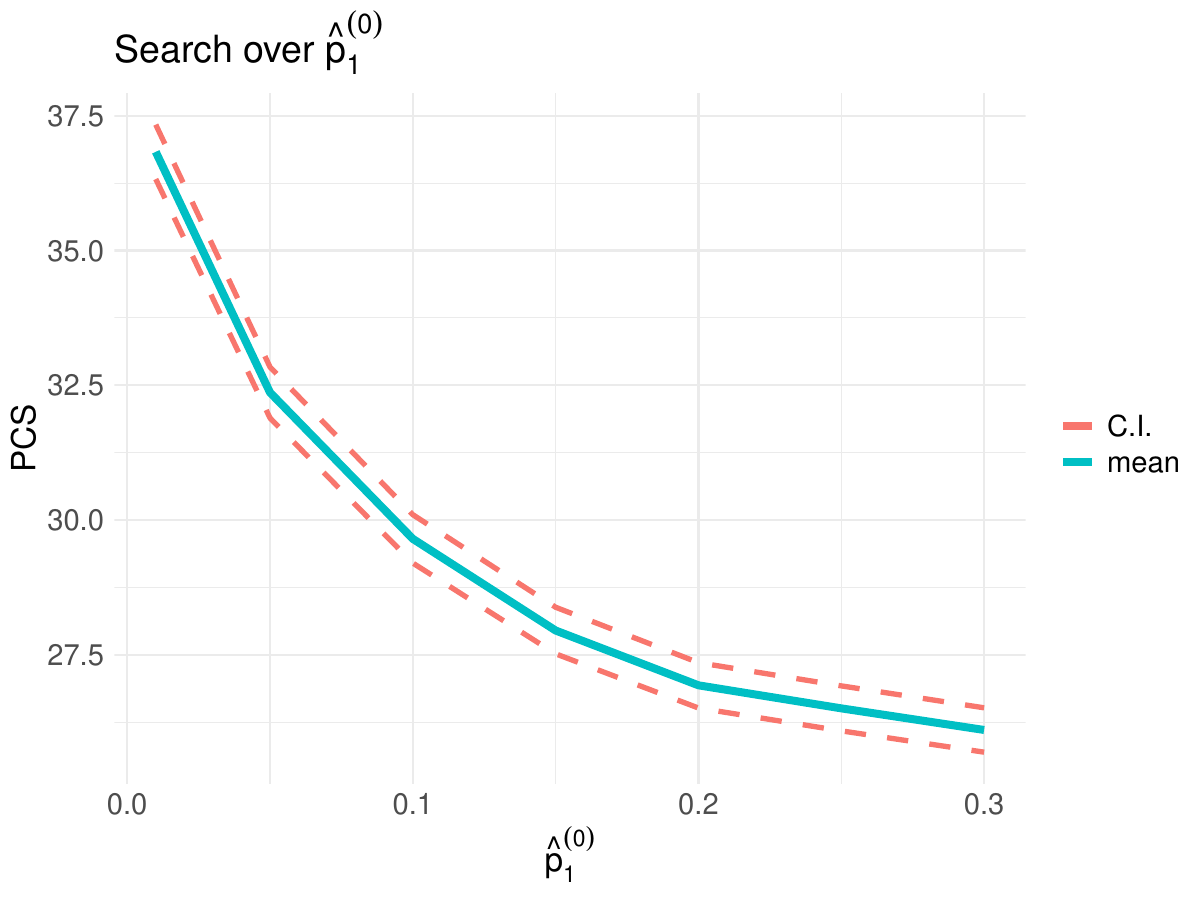}
        \caption{Fix $(\nu,\mu_1,\mu_2,\sigma_1,\sigma_2)=(0.1, 0, 1, 5, 5)$. \label{Subfig: cyclic p1}}
    \end{subfigure}\hfill
    \begin{subfigure}[b]{0.48\linewidth}
        \centering
        \includegraphics[width=\textwidth]{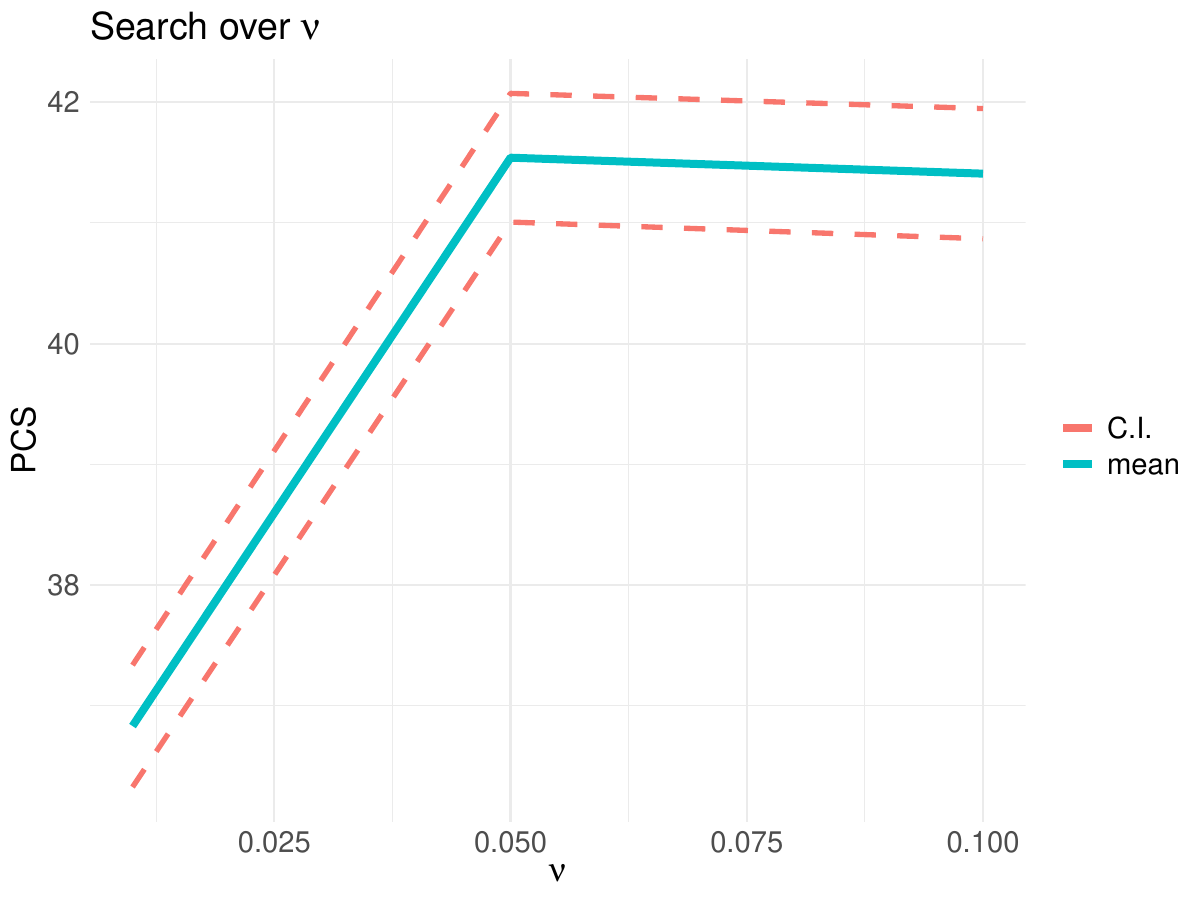}
        \caption{Fix $(\hat{p}_1^{(0)},\mu_1,\mu_2,\sigma_1,\sigma_2)=(0.01, 0, 1, 5, 5)$. \label{Subfig: cyclic nu}}
    \end{subfigure}
    \caption{Cyclic calibration with 90\% C.I.. The estimated PCS in blue and C.I. in red. \label{Fig: calibration details}}
\end{figure}

We fix $\lambda$ in Equation~\eqref{Eqn: conf level score} at 0.2, which gives the optimal significance level $\alpha=0.05$ (see evaluations in Supplementary Materials Table~4). The cyclic algorithm starts from a random initial vector $\omega_0=(\hat{p}_1^{(0)}, \nu,\mu_1,\mu_2,\sigma_1,\sigma_2)=(0.1, 0.1, 0, 1, 5, 5)$. According to the PO-benchmark, Scenario 1 and Scenario 7 correspond to the highest and the lowest proportion of correct selections, and hence these scenarios are chosen for the calibration.  Holding the other parameters fixed and searching along $\hat{p}_1^{(0)}\in\mathfrak{P}$ gives the estimated (based on $10^4$ simulations) geometric mean of the PCS over the selected two scenarios shown in Figure~\ref{Subfig: cyclic p1}. 

The mean PCS is plotted against values of $\hat{p}_1^{(0)}$ with the 95\% C.I. in red dashed lines (Panel (a)). It shows that the lower bound at $\hat{p}_1^{(0)}=0.01$ is higher than the upper bound at the current value $\hat{p}_1^{(0)}=0.1$, and thus $\omega^\ast$ will be updated to $(\hat{p}_1^{(0)}, \nu,\mu_1,\mu_2,\sigma_1,\sigma_2)=(0.01, 0.1, 0, 1, 5, 5)$. The next parameter to be optimised, while keeping the others fixed, is $\nu$, which is shown in Panel (b) of Figure~\ref{Subfig: cyclic nu}. Although updating $\nu$ from 0.1 to 0.05 would increase the mean PCS slightly from 41.41\% to 41.54\%, the upper bound at 0.1 is still higher than the lower bound at 0.05, and thus $\nu$ would stay at the current value 0.1. The algorithm continues for the other parameters in the same manner, which is displayed in Figure~\ref{Fig: cyclic hexagon}.

Each edge of the black hexagon contains the grid of one parameter. The starting point is shown by the red hexagon in~\ref{Subfig: start point} whose axes are the current values of design parameters. The first cycle is shown in blue in Figure~\ref{Subfig: update p1} and~\ref{Subfig: iter1}. The search over $\hat{p}_1^{(0)}$ and $\nu$ results in updating $\hat{p}_1^{(0)}$ and keeping $\nu$ at the current value. The first two edges of the inside hexagon is presented in~\ref{Subfig: update p1} in blue, the old values are shown in the red dashed lines. After one full cycle, the new $\omega^\ast=(\hat{p}_1^{(0)}, \nu,\mu_1,\mu_2,\sigma_1,\sigma_2)=(0.01, 0.1, 1, -1, 1, 1)$ is shown in Figure~\ref{Subfig: iter1} in blue, all parameters except $\nu$ are updated. After the 2nd cycle, only $\hat{p}_1^{(0)}$ and $\nu$ are further updated, and the optimal parameters becomes $\omega^\ast=(\hat{p}_1^{(0)}, \nu,\mu_1,\mu_2,\sigma_1,\sigma_2)=(0.15, 0.01, 1, -1, 1, 1)$, as shown in Figure~\ref{Subfig: iter2} in green. Cycle 3 does not change the value of any parameter, and whence, the algorithm converges. Denote $\omega_\mathrm{POBLRM}=\omega^\ast=(0.15, 0.01, 1, -1, 1, 1)$. 

\begin{figure}[H]
    \centering
    \begin{subfigure}[b]{0.48\linewidth}
        \centering
        \includegraphics[width=\textwidth]{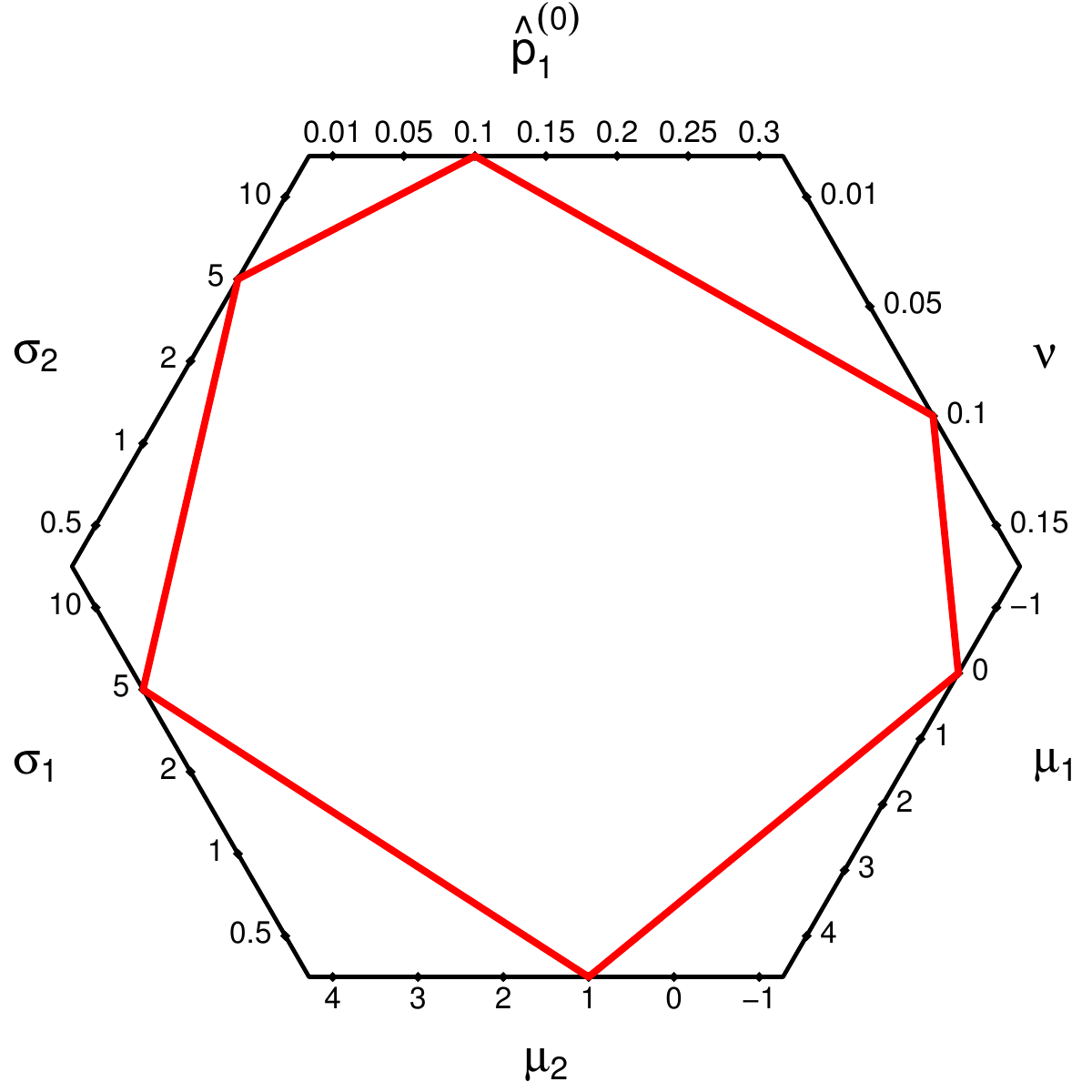}
        \caption{Starting point. \label{Subfig: start point}}
    \end{subfigure} \hfill
    \begin{subfigure}[b]{0.48\linewidth}
        \centering
        \includegraphics[width=\textwidth]{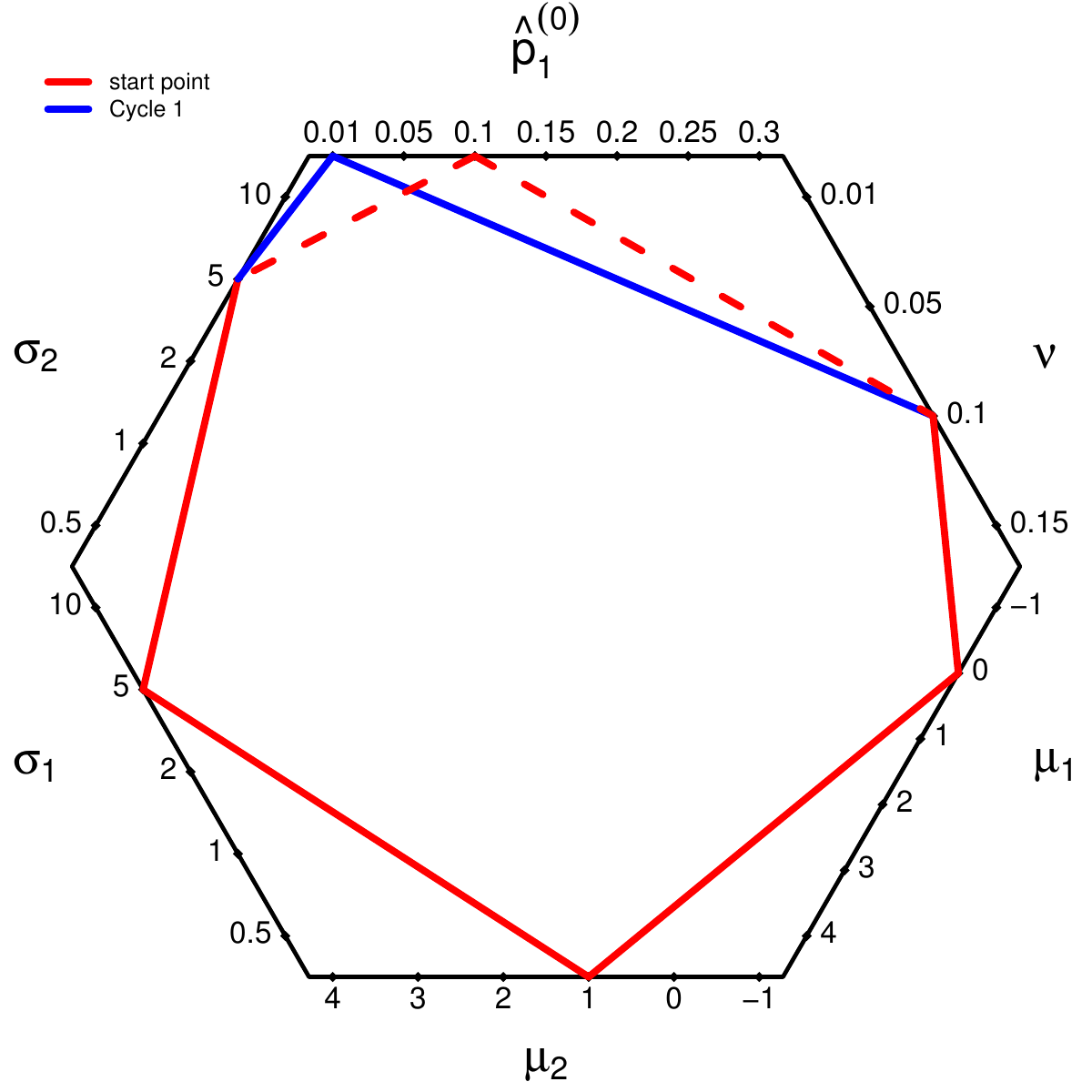}
        \caption{Iteration 1: updating $\hat{p}_1^{(0)}$. \label{Subfig: update p1}}
    \end{subfigure}
\end{figure}

\begin{figure}[H]
\ContinuedFloat
    \centering
    \begin{subfigure}[b]{0.48\linewidth}
        \centering
        \includegraphics[width=\textwidth]{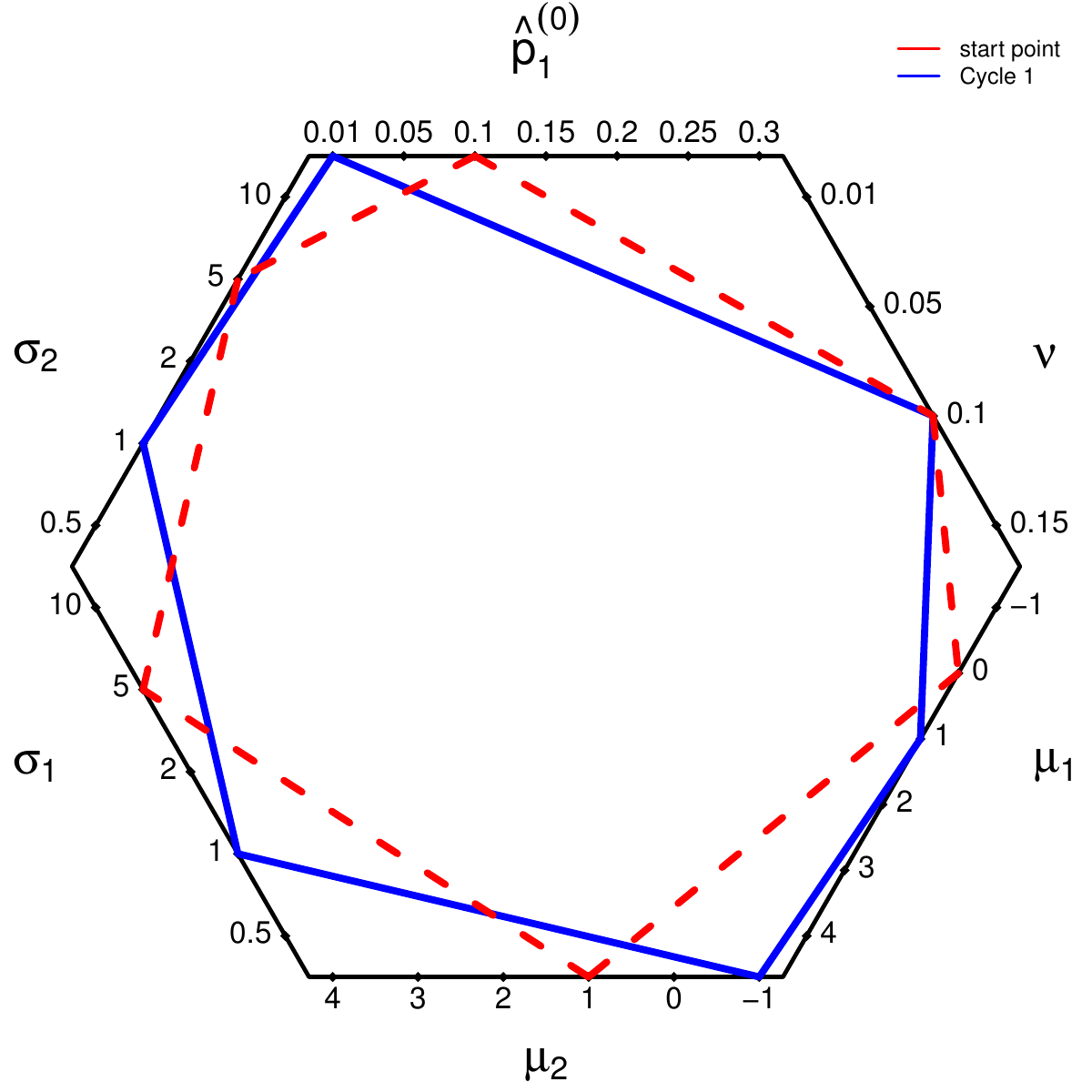}
        \caption{Complete cycle 1. \label{Subfig: iter1}}
    \end{subfigure} \hfill
    \begin{subfigure}[b]{0.48\linewidth}
        \centering
        \includegraphics[width=\textwidth]{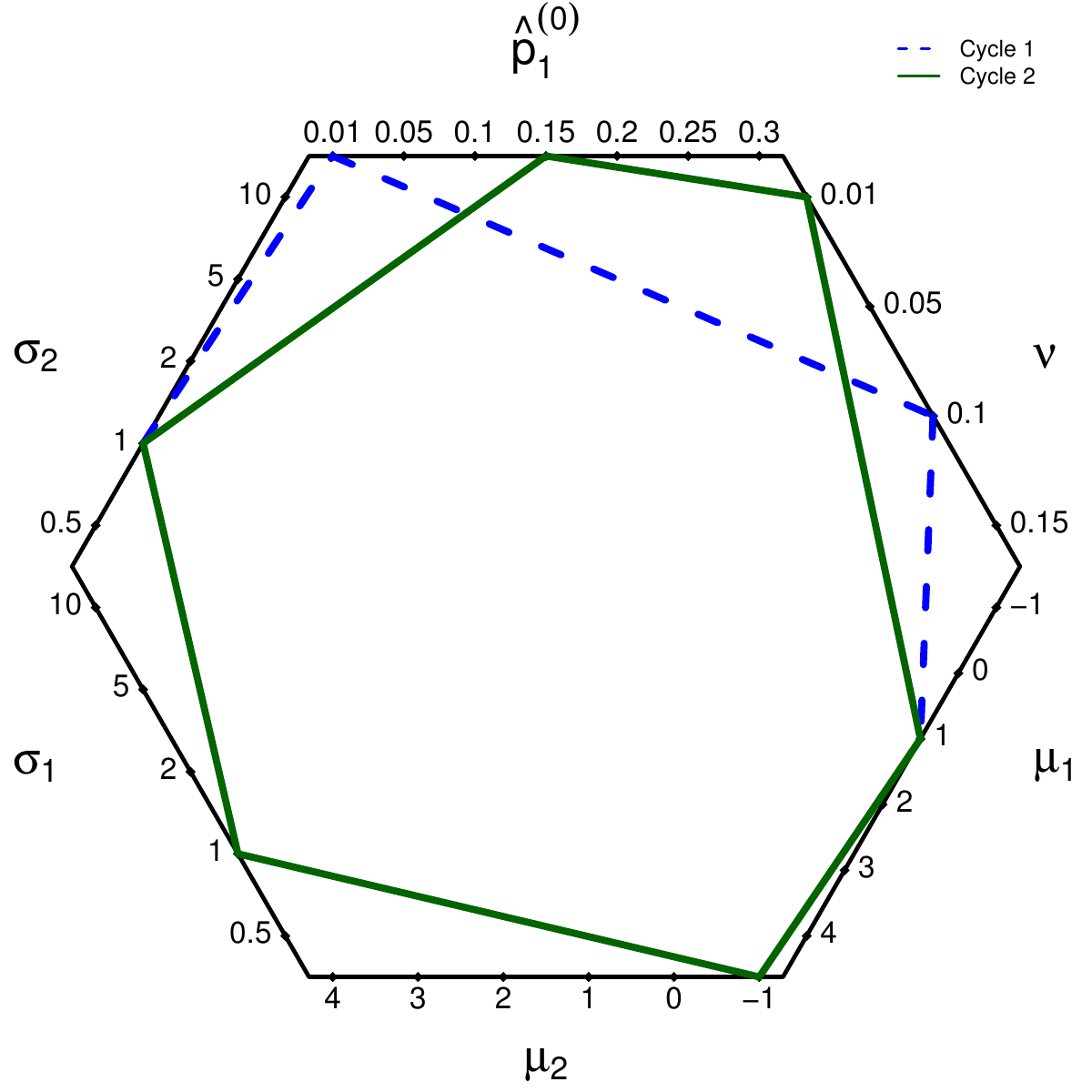}
        \caption{Complete cycle 2. \label{Subfig: iter2}}
    \end{subfigure}
    \caption{Graphical illustration of the cyclic algorithm. \label{Fig: cyclic hexagon}}
\end{figure}

Hence, under the operational priors, the POBLRM uses a normal prior $\theta=(\theta_1,\log\theta_2)^T\sim\mathcal{N}\left(
\begin{pmatrix}
1\\
-1
\end{pmatrix},
\begin{pmatrix}
1 & 0\\
0 & 1
\end{pmatrix}
\right)$,
and the prior toxicity probabilities are equally spaced with the difference $\nu=0.01$ and $\hat{p}_1^{(0)}=0.15$. The matched pseudo prior is $(y_{-1},n_{-1},y_0,n_0)=(0.45, 1.50, 0.57, 1.65)$. We refer to the proposed approach as ``Cyclic2''.

\subsection{Competing calibration approaches}
We compared the proposed calibration approach to the following two
\begin{itemize}
\item[(1)] Grid search using the same grid and all 20 scenarios. Referred to as ``Grid20''
\item[(2)] Cyclic search under all 20 scenarios. Referred to as  ``Cyclic20''
\end{itemize}
Then, under the selected operational priors, the design will be evaluated based on all 20 scenarios. We have also explored choosing  the two scenarios with the highest PCS and two scenarios with the lowest PCS by the benchmark but found that it leads to the same operational prior and similar operating characteristics (see Supplementary Materials). 

Figure~\ref{Fig: compare calibration non-randomised} shows the comparison of the three calibration methods under the POBLRM model. The PCS in all 20 scenarios are similar under all 3 methods, the biggest difference is roughly 4\% under scenario 12, where the PCS under Grid20 is higher than that under the other two approaches. Furthermore, there is no domination of any single approach, all methods can perform the best under some scenarios. 

\begin{figure}[H]
    \centering
    \includegraphics[width=\linewidth]{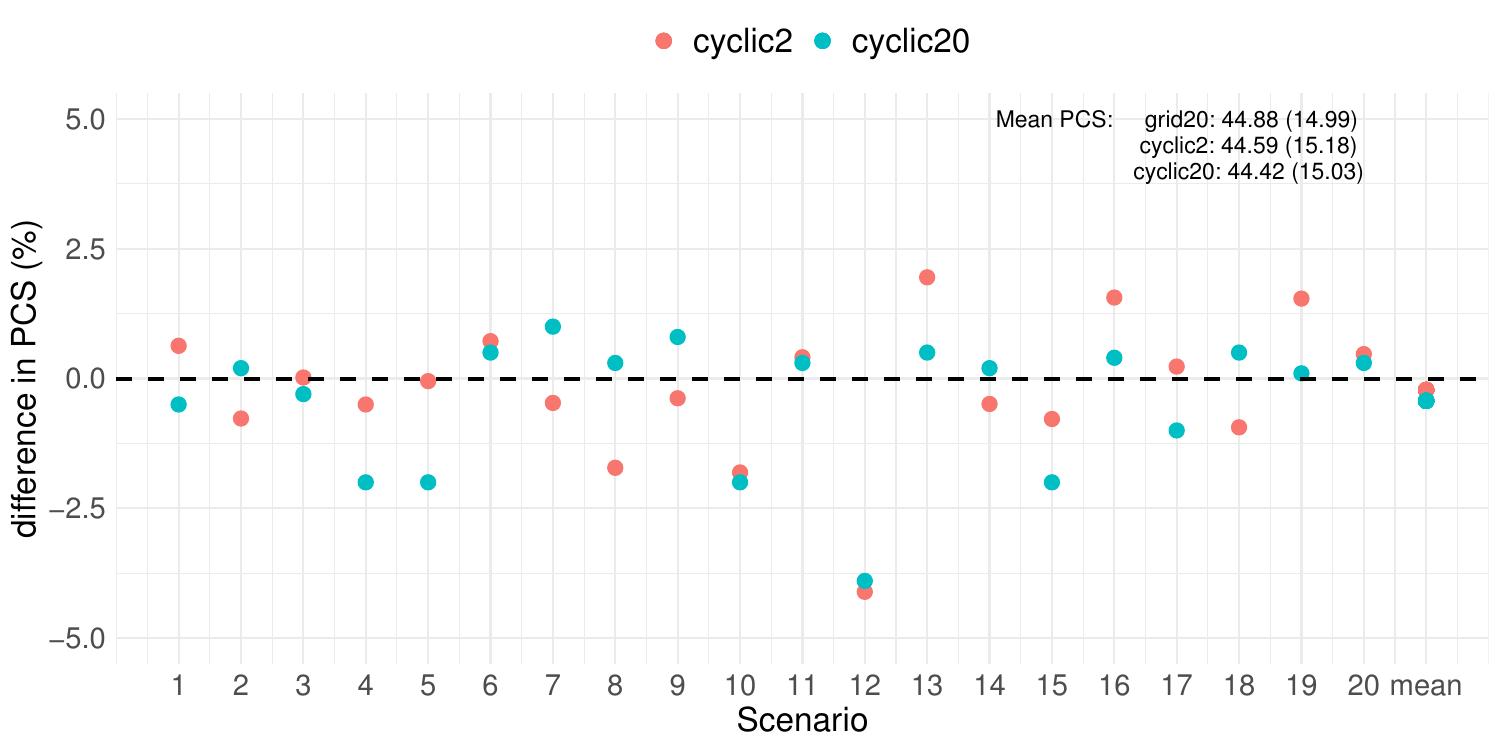}
    \caption{Difference in the PCS for the POBLRM design using different calibration procedures compared to grid search under 20 scenarios with the cyclic under 20 (green) and 2 (red) scenarios. Numbers on the top-right corner are mean (standard error) of the PCS under each calibration procedure. Results are based on $10^4$ simulations. \label{Fig: compare calibration non-randomised}} 
\end{figure}

Regarding computation, each cycle of the cyclic approach evaluates against $|\mathfrak{P}|+|\mathfrak{V}|+2|\mathfrak{M}|+2|\mathfrak{S}|=33$ values of parameters, and the algorithm converges within 3 cycles. Hence, the computation is roughly 100 values under 2 scenarios. Whereas, under grid search with all scenarios, the optimal set of design parameters is $\omega^\ast_\mathrm{grid}=(\hat{p}_1^{(0)}, \nu,\mu_1,\mu_2,\sigma_1,\sigma_2)=(0.15, 0.05, 1, -1, 1, 0.5)$. The computation cost would be $|\mathfrak{P}|\times |\mathfrak{V}|\times |\mathfrak{M}|^2\times |\mathfrak{S}|^2=25200$ under 20 scenarios. The cyclic20 approach selects $\omega^\ast_\mathrm{cyclic20}=(\hat{p}_1^{(0)}, \nu,\mu_1,\mu_2,\sigma_1,\sigma_2)=(0.15, 0.01, 1, -1, 1, 5)$ and converges in 3 cycles. Hence, the computation is 10 times higher than cyclic2.

Overall, evaluating against all 20 scenarios, the mean PCS under Cyclic2, Cyclic20, and Grid20 are 46.8\%, 46.6\% and 47.0\%, respectively. Hence, with less than 0.5\% sacrifice in accuracy, the novel calibration method manages to reduce the computation by more than 2500 folds. The same conclusion holds when calibrating the two other competing designs that we compared our POBLRM to in the next section (see Supplementary Materials).

\subsection{Comparison to competing designs}
\label{Sec: non-randomised comparison}
In this section, we compare the proposed  proposed POBLRM design to two competing model-based design, namely the POCRM introduced in Section~\ref{Subsec: POCRM} and the two-dimensional Bayesian logistic regression model (2BLRM)~\citep{Neuenschwander2015,mozgunov2022dose}. 

Following notations in Section~\ref{Subsec: POCRM}, the POCRM has three design parameters to calibrate, $\hat{p}_1^{(0)}, \nu$, and $\sigma$. The 2BLRM design fits a one-dimensional BLRM on each agent and models their interaction using an odds multiplier. It has 14 design parameters to calibrate, the same 6 parameters as the POBLRM for each agent and two more for the interaction. All parameters are calibrated using the cyclic approach (Algorithm~\ref{Alg: cyclic}) for fair comparisons, all under sceanrios 1 and 7. The operational priors are $\omega_\mathrm{POCRM}=(\hat{p}_1^{(0)},\nu,\sigma)=(0.1, 0.05, 0.5)$ and $\omega_\mathrm{2BLRM}=(\hat{p}_{a1}^{(0)}, \hat{p}_{b1}^{(0)}, \nu_a,\nu_b, \mu_{a1}, \mu_{a2}, \mu_{b1}, \mu_{b2},\mu_\zeta, \sigma_{a1}, \sigma_{a2}, \sigma_{b1}, \sigma_{b2},\sigma_\zeta)=(0.30, 0.20, 0.15, 0.15, -1, 0, 2, 0, 0, 1, 1, 1)$.

The operating characteristics of the three designs and the PO-benchmark are compared in Figure~\ref{Fig: nonrandomised} with further details in Table~7 in supplementary materials. The mean PCS with 95\% C.I.s are given in the top-right. On average, the three designs perform similarly, all with mean PCS around 44\%. However, the 2BLRM has a less balanced performance such that it can have particularly high PCS in easy scenarios, such as 1 and 2, but very low PCS in other scenarios, such as 4 and 7. Additionally, when there are more than one MTCs, (scenario 11 - 20), methods adopting partial ordering (POCRM and POBLRM) often have more balanced selection of the MTCs, whereas, 2BLRM often prefer one MTC over the others.

Another considered metric is the probability of overdosing, i.e. recommend a combination with higher toxicity than the TTL, which are plotted in Figure~\ref{Fig: nonrandomised overdose}. For all three designs, no over-toxic control rule is implemented to better understand the behaviour of the models. Hence, under scenario 1, the overdosing probability is 100\%. Under scenario 10, there is no overly toxic dose. Under most scenarios, the three models either all have similar overdosing probabilities or the 2BLRM has a much higher probability than the others. The mean probability under 2BLRM is also 7\% higher than the other two, which are both around 24\%. Note the scenarios where 2BLRM has particularly high overdosing, such as 3, 15, 18, are what defined as ``antagonistic" scenarios in~\citet{widmer2023principled}. That means, the toxicity of the dose-combination is lower than the two drugs acting independently. For example, under scenario 3, $d_5$ is less toxic than the sum of the DLT probabilities of the second level of each drug. In the paper, they argued that 2BLRM is not suitable for such scenarios.

\begin{figure}[H]
    \centering
    \begin{subfigure}[b]{0.9\linewidth}
        \centering
        \includegraphics[width=\linewidth]{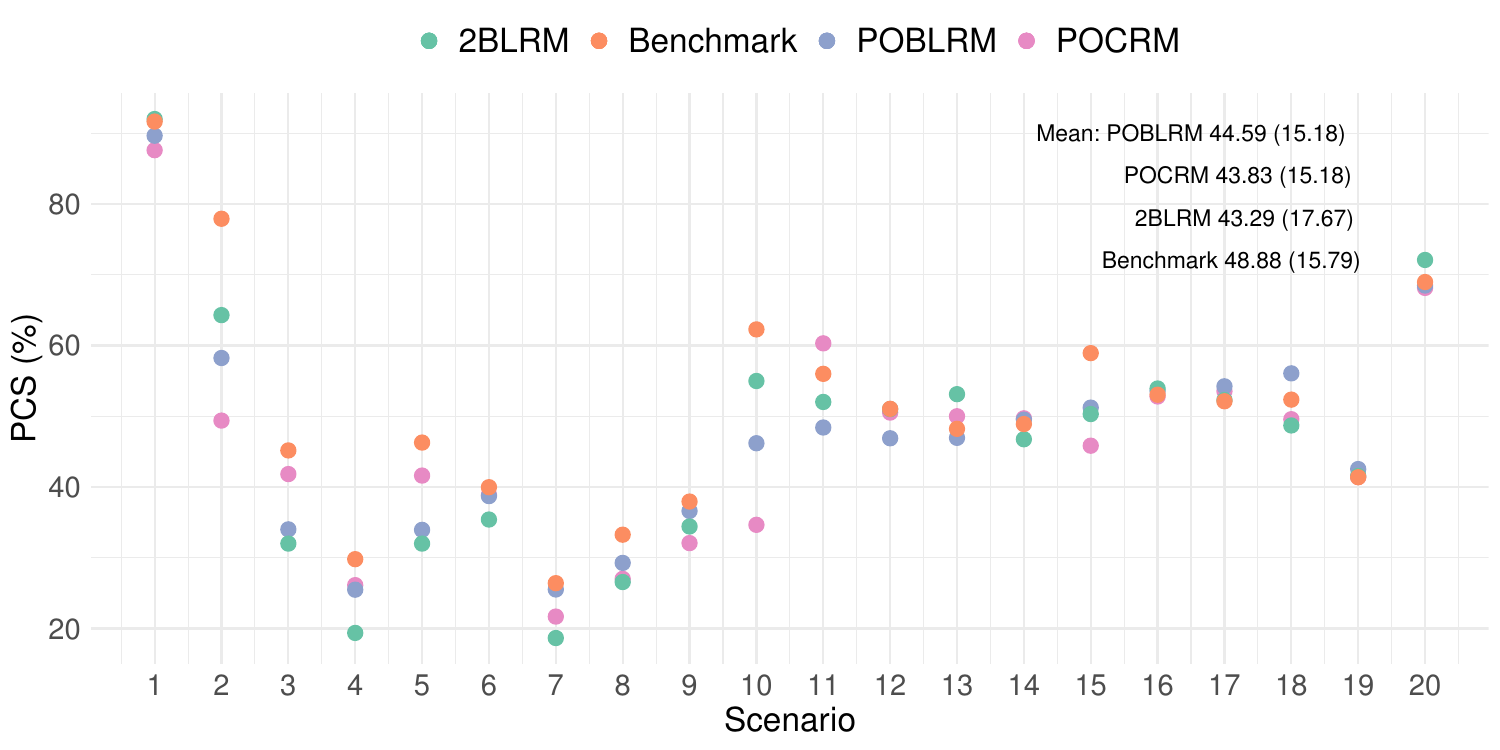}
        \caption{PCS for the POCRM (purple), POBLRM (blue), 2BLRM (green), and PO-benchmark (orange) using the cyclic2 calibration procedure. Numbers on the top-right corner are mean (standard error) of the PCS under each model. Results are based on $10^4$ simulations. \label{Fig: nonrandomised}}
    \end{subfigure} 
    \begin{subfigure}[b]{0.9\linewidth}
        \centering
        \includegraphics[width=\linewidth]{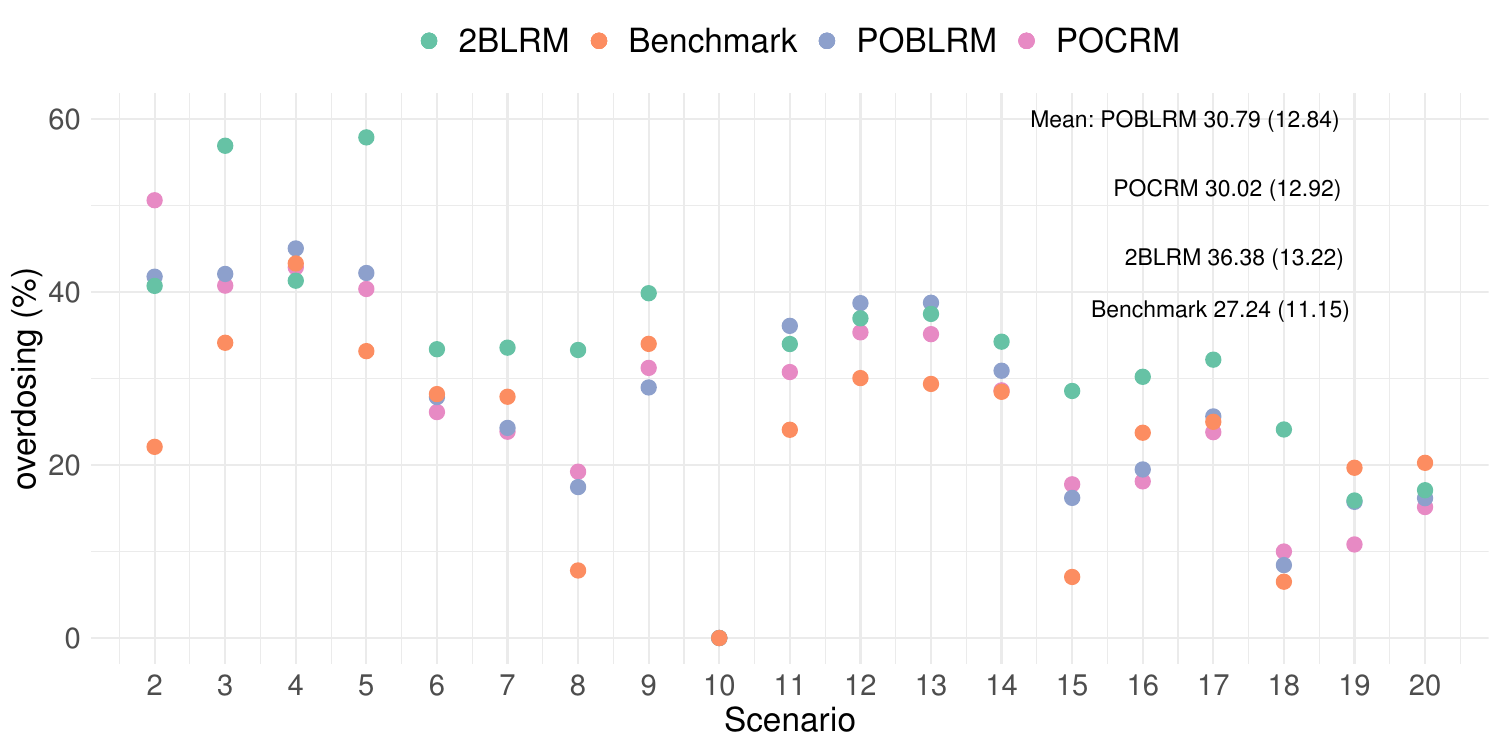}
        \caption{The probability of overdosing under the POCRM (purple), POBLRM (blue), 2BLRM (green), and PO-benchmark (orange). The design parameters are calibrated via the cyclic2 procedure. Numbers on the top-right corner are mean (standard error) of the overdosing probability under each model. Results are based on $10^4$ simulations. \label{Fig: nonrandomised overdose}}
    \end{subfigure}
    \caption{Comparing the operational characteristics of the three designs. \label{Fig: nonrandomised compare}}
\end{figure}

Finally, regarding running times, for 45 patients enrolled in 15 cohorts, POCRM is the fastest, which takes on average only 0.6 second for one simulation for one scenario. The POBLRM takes slightly longer, 1.36 seconds for each simulation. However, the running time for the 2BLRM is dramatically longer, taking 95.42 seconds for each trial. The relative time for POCRM : POBLRM : 2BLRM is 1 : 2.27 : 170.39.

\subsection{Numerical evaluation in the randomised setting}
\label{Sec: randomised}
The second motivating example was the COVID-19 trial that needs to include a control group in the study and targeting the risk of the additional toxicity compared to the risk of the SoC.  In this section, the POBLRM and POCRM are applied to the setting where patients are randomised between SoC and a grid of treatment dose-combinations. 

For each cohort of $m=m_1+m_2$ patients, $m_2$ are randomised to the control group, and the rest $m_1$ are allocated to the current MTC. The design is detailed in Algorithm~\ref{Alg: POBLRM} under \verb|randomised=TRUE|. If the estimated toxicity of SoC is closest to the TTL, the $m_1$ patients will also be allocated to SoC. Hence, at least $\frac{m_2}{m_1+m_2}N$ patients are at the control group. 

The allocation ratio of $m_1:m_2=3:1$ is used. All the design parameters are calibrated using the cyclic calibration, which gives $\omega_\mathrm{POBLRM}=(\hat{p}_1^{(0)}, \nu, \mu_1, \mu_2, \sigma_1,\sigma_2)=(0.05, 0.07, 1, -1, 1, 1)$ and $\omega_\mathrm{POCRM}=(\hat{p}_1^{(0)}, \nu, \sigma)=(0.05, 0.07, 0.5)$. Both algorithms converge within 2 cycles. Under the POBLRM, the matched pseudo prior is $(y_{-1},n_{-1}, y_0, n_0)=(0.37, 1.37, 2.86, 4.54)$. 

We will compare the operating characteristics to POCRM under the same allocation rule. The same 20 scenarios in Table~\ref{Tab: example scenarios} are used while the toxicities of the SoC are given in Table~\ref{Tab: SoC toxicity}. 
\begin{table}[H]
    \centering
    \begin{tabular}{l cccccccccc}
    \hline
    Scenario & 1 & 2 & 3 & 4 & 5 & 6 & 7 & 8 & 9 & 10\\ \cline{2-11}
    Toxicity & 0.30 & 0.15 & 0.10 & 0.05 & 0.05 & 0.05 & 0.05 & 0.01 & 0.01 & 0.01\\ \hline
    Scenario & 11 & 12 & 13 & 14 & 15 & 16 & 17 & 18 & 19 & 20\\ \cline{2-11}
    Toxicity & 0.05 & 0.05 & 0.05 & 0.05 & 0.01 & 0.01 & 0.05 & 0.01 & 0.01 & 0.05\\ \hline
    \end{tabular}
    \caption{Toxicity probability of the Standard of care ($d_0$) under the 20 scenarios given in Table~\ref{Tab: example scenarios}.\label{Tab: SoC toxicity}}
\end{table}  

The probabilities of selecting each combination are summarised in Table~\ref{Tab: randomised}. The biggest difference lies in scenario 1, where all combinations are overly toxic and the SoC should be selected. The more flexible BLRM can indeed select the SoC with 78\% probability, whereas, with CRM, the SoC and $d_1$ are selected with similar frequencies. Under scenario 10, when all combinations are save, the POBLRM can indeed select the highest combination with highest frequency around 30\%. Wheares, the CRM selects lower combinations more frequently due to the shape of the power model, and it only selects $d_9$ approximately 6\% of the times. Among the other scenarios, these two models behave similarly. 

The mean PCS under the POBLRM and POCRM are 45\% and 38\%, respectively. This difference is as expected, since the 2-parameter model is capable of accurately estimating at 2 combinations (SoC and the MTC), whereas, the fitting under the 1-parameter model is a compromise between the two combinations, giving inaccurate estimate at either of them.

\begin{table}[H]
    \centering
    \begin{tabular}{ll |cccccccccc| r}
        \hline
        \multicolumn{2}{c|}{Scenario} & \multicolumn{10}{c|}{Proportion (\%)} & PCS\\
        && SoC & 1 & 2 & 3 & 4 & 5 & 6 & 7 & 8 & 9 & \\ \cline{3-12}
        & POBLRM & \textbf{77.6} & 19.4 & 1.8 & 0.1 & 1.2 & 0 & 0 & 0.0 & 0 & 0  & 77.6\\ 
        \multirow{-2}{*}{1} & POCRM & \textbf{45.5} & 41.8 & 7.5 & 0.3 & 4.7 & 0.0 & 0 & 0.3 & 0 & 0 & 45.5\\ \hline

        & POBLRM & 10.3 & \textbf{44.3} & 20.1 & 2.3 & 20.3 & 0.6 & 0.0 & 2.2 & 0.0 & 0  & 44.3\\ 
        \multirow{-2}{*}{2} & POCRM & 4.0 & \textbf{42.7} & 23.5 & 3.0 & 23.2 & 0.6 & 0.0 & 3.0 & 0.0 & 0 & 42.7\\ \hline

        & POBLRM & 0.6 & 14.5 & \textbf{41.6} & 8.5 & 29.0 & 1.3 & 0.1 & 4.3 & 0.1 & 0.0 & 41.6\\ 
        \multirow{-2}{*}{3} & POCRM & 0.1 & 15.5 & \textbf{44.3} & 7.3 & 28.0 & 1.1 & 0.0 & 3.8 & 0.0 & 0 & 44.3\\ \hline

        & POBLRM & 0 & 0.0 & 12.5 & \textbf{25.4} & 17.3 & 23.6 & 2.7 & 16.5 & 1.9 & 0.0 & 25.4\\
        \multirow{-2}{*}{4} & POCRM & 0 & 0.1 & 14.7 & \textbf{27.6} & 19.7 & 20.9 & 1.2 & 15.1 & 0.7 & 0 & 27.6\\ \hline

        & POBLRM & 0.0 & 7.4 & 30.2 & 5.8 & \textbf{43.1} & 2.7 & 0.1 & 10.8 & 0.0 & 0 & 43.1\\
        \multirow{-2}{*}{5} & POCRM & 0.1 & 14.6 & 28.2 & 4.1 & \textbf{44.1} & 1.3 & 0.0 & 7.7 & 0.0 & 0 & 44.1\\ \hline

        & POBLRM & 0 & 0.0 & 3.5 & 18.7 & 13.1 & \textbf{38.7} & 6.5 & 16.5 & 2.9 & 0.0 & 38.7\\
        \multirow{-2}{*}{6} & POCRM & 0 & 0.0 & 3.8 & 19.1 & 15.1 & \textbf{40.5} & 5.4 & 14.3 & 1.7 & 0 & 40.5\\ \hline

        & POBLRM & 0 & 0.0 & 5.5 & 14.7 & 8.4 & 25.8 & \textbf{23.2} & 18.4 & 3.7 & 0.3 & 23.2\\
        \multirow{-2}{*}{7} & POCRM & 0 & 0.1 & 7.8 & 18.5 & 12.0 & 27.3 & \textbf{15.6} & 16.3 & 2.3 & 0.0 & 15.6\\ \hline

        & POBLRM & 0 & 0 & 0.4 & 24.1 & 0.6 & 24.8 & 10.5 & \textbf{30.9} & 8.7 & 0.1 & 30.9\\
        \multirow{-2}{*}{8} & POCRM & 0 & 0 & 1.2 & 22.7 & 1.7 & 30.8 & 9.0 & \textbf{27.6} & 6.9 & 0 & 27.6\\ \hline

        & POBLRM & 0 & 0 & 0.8 & 24.7 & 0.3 & 19.8 & 5.1 & 9.5 & \textbf{39.7} & 0.2 & 39.7\\
        \multirow{-2}{*}{9} & POCRM & 0 & 0 & 2.2 & 22.9 & 1.0 & 28.7 & 3.3 & 12.4 & \textbf{29.6} & 0.0 & 29.6\\ \hline

        & POBLRM & 0 & 0 & 0.5 & 7.8 & 0.4 & 4.6 & 24.3 & 7.5 & 25.0 & \textbf{30.0} & 30.0\\
        \multirow{-2}{*}{10} & POCRM & 0 & 0 & 1.8 & 14.4 & 1.8 & 15.8 & 22.7 & 14.3 & 22.9 & \textbf{6.4} & 6.4\\ \hline

        & POBLRM & 0.0 & 4.3 & \textbf{30.6} & 8.4 & \textbf{27.9} & 10.3 & 0.3 & 17.0 & 1.1 & 0 & 58.5\\
        \multirow{-2}{*}{11} & POCRM & 0.0 & 6.5 & \textbf{32.4} & 6.7 & \textbf{31.6} & 7.2 & 0.1 & 15.2 & 0.4 & 0 & 63.9\\ \hline

        & POBLRM & 0.0 & 0.3 & \textbf{23.6} & 13.0 & 12.5 & 19.4 & 1.8 & \textbf{26.2} & 3.0 & 0.1 & 49.8\\
        \multirow{-2}{*}{12} & POCRM & 0 & 0.5 & \textbf{29.2} & 10.9 & 15.2 & 15.2 & 0.6 & \textbf{27.3} & 1.0 & 0 & 56.5\\ \hline

        & POBLRM & 0 & 0.2 & 12.6 & \textbf{26.1} & \textbf{23.8} & 20.0 & 2.5 & 13.0 & 1.8 & 0.0 & 49.9\\
        \multirow{-2}{*}{13} & POCRM & 0 & 0.5 & 15.1 & \textbf{28.0} & \textbf{28.7} & 15.2 & 1.0 & 10.9 & 0.7 & 0 & 56.6\\ \hline

        & POBLRM & 0.0 & 0.1 & 8.7 & \textbf{20.5} & 12.9 & \textbf{28.8} & 9.8 & 16.0 & 3.1 & 0.1 & 49.3\\
        \multirow{-2}{*}{14} & POCRM & 0 & 0.1 & 11.8 & \textbf{22.7} & 14.9 & \textbf{27.8} & 6.0 & 14.8 & 1.8 & 0 & 50.5\\ \hline

        & POBLRM & 0 & 0 & 1.5 & \textbf{25.6} & 1.5 & 26.5 & 9.0 & \textbf{27.0} & 8.9 & 0.0 & 52.6\\
        \multirow{-2}{*}{15} & POCRM & 0 & 0 & 3.8 & \textbf{24.0} & 3.6 & 30.9 & 6.8 & \textbf{24.0} & 6.9 & 0 & 48.0\\ \hline

        & POBLRM & 0 & 0 & 0.5 & \textbf{26.2} & 0.2 & 12.1 & 17.5 & 9.7 & \textbf{31.8} & 2 & 57.9\\
        \multirow{-2}{*}{16} & POCRM & 0 & 0 & 1.4 & \textbf{25.9} & 0.7 & 22.0 & 11.0 & 13.5 & \textbf{25.1} & 0.4 & 50.9\\ \hline

        & POBLRM & 0 & 0.1 & 12.1 & 16.5 & 9.8 & \textbf{30.9} & 2.7 & \textbf{23.0} & 4.9 & 0.0 & 53.9\\
        \multirow{-2}{*}{17} & POCRM &   0 & 0.1 & 15.4 & 15.3 & 12.0 & \textbf{28.9} & 1.7 & \textbf{23.4} & 3.3 & 0 & 52.3\\ \hline

        & POBLRM & 0 & 0 & 0.9 & 15.4 & 1.2 & 15.0 & \textbf{31.4} & \textbf{27.2} & 8.6 & 0.4 & 58.6\\
        \multirow{-2}{*}{18} & POCRM &  0 & 0 & 2.6 & 18.6 & 3.1 & 24.4 & \textbf{20.9} & \textbf{23.9} & 6.4 & 0.1 & 44.8\\ \hline

        & POBLRM & 0 & 0 & 1.9 & 13.0 & 1.2 & 11.4 & \textbf{25.2} & 13.0 & \textbf{24.7} & 9.6 & 49.9\\
        \multirow{-2}{*}{19} & POCRM & 0 & 0 & 5.9 & 16.7 & 3.2 & 25.5 & \textbf{14.9} & 17.4 & \textbf{15.3} & 1.2 & 30.2\\ \hline

        & POBLRM & 0 & 0.1 & 8.8 & \textbf{21.4} & 8.1 & \textbf{28.2} & 4.7 & \textbf{19.0} & 9.7 & 0.1 & 68.9\\
        \multirow{-2}{*}{20} & POCRM & 0 & 0.1 & 10.7 & \textbf{20.5} & 10.8 & \textbf{29.2} & 3.2 & \textbf{20.0} & 5.5 & 0.0 & 69.7\\ \hline

        && \multicolumn{10}{c|}{POBLRM} & \textbf{45.11}\\
        \multirow{-2}{*}{\textbf{Mean}} && \multicolumn{10}{c|}{POCRM} & \textbf{38.16}\\\hline 
    \end{tabular}
    \caption{Randomised between SoC and treatment dose-combinations. \label{Tab: randomised}}
\end{table}

\section{Discussion}
A novel two-parameter POBLRM has been proposed for dual-agent combination-escalation trials. Compared to POCRM, it is more flexible, and suitable in settings where questions beyond MTC are of interest, such as large combination-schedule grids and randomisation of patients between the SoC and treatment. It was shown to give the same or higher PCS in the non-randomised setting, and in a noticeably higher PCS in the randomised setting.

A link between two commonly used priors for BLRM models, normal and pseudo, has been established by minimising the KL-divergence between the combination-toxicity curves. This allowed to increase the computation efficiency and brought an easier interpretation of prior hyperparameters. The pseudo parameters $n_{-1}$ and $n_0$ are the effective sample size corresponding to the prior, giving a direct measure of the amount of information contained.

To accommodate the problem of expensive computation for the proposed and dose-escalation design more generally, a more efficient (than grid search) calibration method, the cyclic algorithm, was proposed. Its computation cost grows additively with the number of design parameters, whereas, the conventional grid search has multiplicative computation. Moreover, convergence to local maximum can be guaranteed by this cyclic approach. In the worst-case scenario, the algorithm converges after exhausted all values of the design parameters, leading to the same computation as the grid search. However, empirically, fast convergence has been observed. For all the simulation studies we investigated, the algorithm converges after 2 or 3 cycles, reducing the computation by more than 5000 folds compared to the grid search. 

Calibration based on a small subset of scenarios has been examined. Simulations shown that the subset should be chosen based on the difficulty of scenarios, measured by PO-benchmark. Based on only the simplest and hardest scenarios, the selected operational prior gives very similar mean PCS across all scenarios compared to calibration using the full set.  

For further studies, theoretical results of POBLRM are worth discussing, such as consistency and coherence. Consistency refers to the property that the probability of selecting the true MTC goes to 1 as the sample size grows to infinity, and coherence considers the cohort-to-cohort behaviour of the trial. If a toxicity is observed, the probability of escalation should go to zero, and conversely, if a non-toxicity is observed, the probability of deescalation should go to zero. These properties are important for better understandings of the design.


\section*{Acknowledgements}

This report is independent research supported by the National Institute for Health Research (NIHR300576). The views expressed are those of the authors and not necessarily those of the NHS, the National Institute for Health Research or the Department of Health and Social Care (DHSC). PM also received funding from UK Medical Research Council (MC UU 00040/03, MC UU 00002/19). For the purpose of open access, the author has applied a Creative Commons Attribution (CC BY) licence to any Author Accepted Manuscript version arising. WC receives the Gates Cambridge Scholarship for her PhD in Biostatistics.\vspace*{-8pt}

\clearpage
\bibliography{References}

\end{document}